\documentclass[pdflatex]{sn-jnl}

\usepackage{graphicx}%
\usepackage{multirow}%
\usepackage{amsmath,amssymb,amsfonts}%
\usepackage{amsthm}%
\usepackage{mathrsfs}%
\usepackage[title]{appendix}%
\usepackage{xcolor}%
\usepackage{textcomp}%
\usepackage{manyfoot}%
\usepackage{booktabs}%
\usepackage{algorithm}%
\usepackage{algorithmicx}%
\usepackage{algpseudocode}%
\usepackage{listings}%

\usepackage{authblk}
\usepackage{natbib}       
\usepackage{hyperref}
\usepackage{cleveref}
\usepackage{mathtools}    
\usepackage{bbm}          
\usepackage[T1]{fontenc}  
\usepackage[utf8]{inputenc}
\usepackage{tabularx}     
\usepackage{url}

\usepackage{geometry}
\usepackage{etoolbox}
\crefname{appendix}{Appendix}{Appendices}
\Crefname{appendix}{Appendix}{Appendices}
\AtBeginEnvironment{appendices}{%
  \crefalias{section}{appendix}%
}

\theoremstyle{thmstyleone}%
\theoremstyle{thmstyletwo}%
\newtheorem{remark}{Remark}%
\theoremstyle{thmstylethree}%
\newtheorem{definition}{Definition}%

\begin{document}

\title[Readout Representation: Redefining Neural Codes by Input Recovery]{Readout Representation: Redefining Neural Codes by Input Recovery}

\author[1]{Shunsuke Onoo}

\author[1,2]{Yoshihiro Nagano}

\author[1, 2, 3]{Yukiyasu Kamitani}

\affil[1]{Graduate School of Informatics, Kyoto University, Kyoto, Japan}
\affil[2]{ATR Computational Neuroscience Laboratories, Kyoto, Japan}
\affil[3]{Guardian Robot Project, RIKEN, Kyoto, Japan}

\abstract{Sensory representation is typically understood through a hierarchical-causal framework where progressively abstract features are extracted sequentially. However, this causal view fails to explain misrepresentation, a phenomenon better handled by an informational view based on decodable content. This creates a tension: how does a system that abstracts away details still preserve the fine-grained information needed for downstream functions? We propose \textit{readout representation} to resolve this, defining representation by the information recoverable from features rather than their causal origin.
Empirically, we show that inputs can be accurately reconstructed even from heavily perturbed mid-level features, demonstrating that a single input corresponds to a broad, redundant region of feature space, challenging the causal mapping perspective.
To quantify this property, we introduce \textit{representation size}, a metric linked to model robustness and representational redundancy. Our framework offers a new lens for analyzing how both biological and artificial neural systems learn complex features while maintaining robust, information-rich representations of the world.}

\maketitle

\section{Introduction}
The dominant view of neural sensory representation is rooted in a hierarchical-causal framework, where representations are the causal outcome of a stimulus processed through layers that extract progressively abstract features \citep{dicarlo_how_2012,kriegeskorte_representational_2008}. This model, aligned with causal theories in philosophy \citep{Fodor1987-FODPTP}, has spurred powerful analytical tools \citep{kornblith_similarity_2019,seung_manifold_2000}. However, its strict causal foundation fails to explain misrepresentation—phenomena like illusions or mental imagery, where representational content is decoupled from a direct sensory cause. Alternative philosophical theories address this gap: informational accounts define representation by the information a state carries \citep{Dretske1981-DREKAT}, while teleological accounts define it by its proper function for a downstream consumer \citep{Millikan1989-MILB}. These non-causal views are empirically supported by neural decoding studies, which show that subjective content like dreams and imagery can be read out from brain activity using decoders trained with stimulus-induced perception \citep{horikawa2013neural,cheng_reconstructing_2023,kamitani2025visual}.

These competing perspectives create a central tension: how can a system designed for hierarchical abstraction---which supposedly discards details---simultaneously preserve the fine-grained information required for downstream functions? Both theoretical and empirical work have shown that fine-grained information remains recoverable even from higher-level representations with large receptive fields \citep{zhang1999neuronal,majima2017position}. In deep neural networks, detailed visual appearances can be reconstructed from upper layers \citep{mahendran_understanding_2014}. These findings confirm that abstraction and detail retention are not mutually exclusive, suggesting that population codes can support both simultaneously.

\begin{figure}[t]
    \centering
    \includegraphics[
        width=1\linewidth
    ]{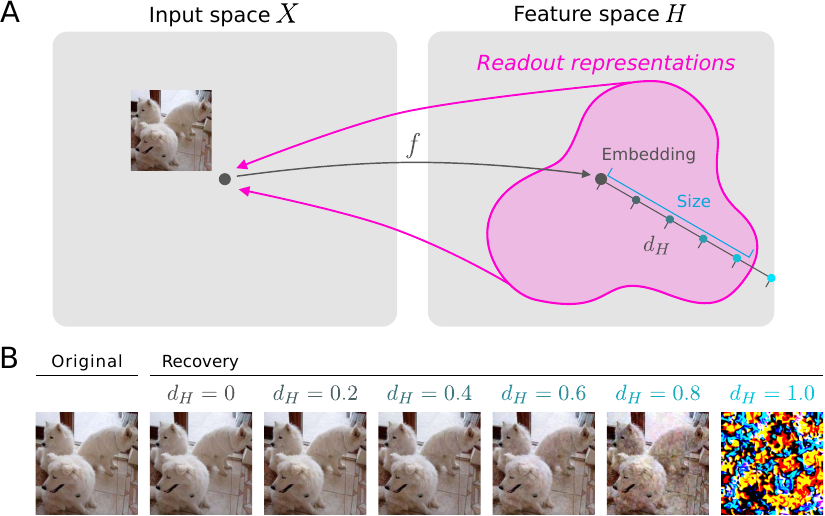}
    \caption{
        Concept of \textit{readout representation} and images reconstructed from perturbed features. \textbf{A}: Concept of readout representation. Traditionally, a neural representation is defined by causal relationship—by the input that elicits a representation. In contrast, we propose to define representations by the information recoverable from neural features. Under this framework, representation is decoupled from its causal origin. The representation of a single input can be a region in feature space, consisting of all points from which the input information can be readout. \textbf{B}: Input images can be recovered from heavily perturbed features. From left to right, the original image, the image recovered from the original feature, and the images recovered from perturbed features. On recovery images, indicate the correlation distance between the original feature and the perturbed features. Reconstruction was performed using the conv3\_1 layer in VGG19. This finding shows that an input image is represented by a broad area in the representational space, supporting the readout view. We provide full results in \Cref{sec:results_vision} and \Cref{sec:results_language}.
        }
    \label{fig:fig1_concept}
\end{figure}

To formally reconcile these observations, we introduce \textit{readout representation}, a framework that operationalizes insights from informational and teleological theories. We define a representation not by its causal origin, but as the set of all neural features from which a specific signal can be functionally recovered (Figure \hyperref[fig:fig1_concept]{1A}). This approach provides a concrete, quantifiable method for analyzing how information is preserved throughout a system's hierarchy, moving beyond abstract philosophical distinctions. It offers a practical lens for studying how both biological and artificial systems learn abstract features while maintaining robust, information-rich codes.

To explore our framework's implications, we used deep neural networks as a fully observable testbed. Our investigation revealed a striking phenomenon: a single input corresponds to a vast and continuous region of feature space from which it can be recovered. To systematically characterize this discovery, we probed the boundaries of these readout representations by applying feature inversion \citep{mahendran_understanding_2014} to deliberately perturbed features. We found these representational sets to be surprisingly large across both vision and language models, with inputs remaining recoverable even from perturbed features displaced far from their canonical values (Figure \hyperref[fig:fig1_concept]{1B}). To quantify this property, we introduce \textit{representation size}, a metric capturing this robustness that correlates with representational redundancy and model performance. Experiments with simplified models further confirm that this is a general principle of neural representation, not an artifact of deep networks.

Our key contributions are as follows:
\begin{itemize}
    \item \textbf{Conceptual framework}: We introduce \textit{readout representation}, a framework that defines representation by functional recoverability rather than causal origin, resolving the tension between hierarchical processing and information preservation.
    \item \textbf{Empirical validation}: We demonstrate across diverse models that input information is recoverable from a broad range of perturbed features, establishing the generality of our framework.
    \item \textbf{Quantitative measure}: We propose \textit{representation size} as a novel metric that captures the extent of this recoverable feature space, linking it to representational redundancy and model robustness.
\end{itemize}

\section{Related work}

\textbf{Philosophical theories of representation.} Classical accounts often define neural representations in terms of their causal origin, where features are taken to represent the stimuli that produced them \citep{Fodor1987-FODPTP}. This view struggles with misrepresentation, such as illusions or imagery, where representational content diverges from external causes. Informational theories instead define representation by the information carried by a state \citep{Dretske1981-DREKAT}, while teleosemantic accounts emphasize proper function for downstream consumers \citep{Millikan1989-MILB}. Our work operationalizes these non-causal perspectives by defining the representation in terms of recoverability.

\textbf{Hierarchical-causal views in neuroscience and machine learning.} Much empirical work interprets neural representations through progressive abstraction of input stimuli. This has motivated concepts such as neural manifolds \citep{seung_manifold_2000,dicarlo_how_2012,poole_exponential_2016,sorscher_neural_2022,chung_classification_2018}, representational similarity analysis \citep{kriegeskorte_representational_2008,kornblith_similarity_2019,huh_platonic_2024}, and information bottleneck theory \citep{tishby2015deep,shwartz2017opening}. Hierarchical receptive-field analyses in deep networks further reveal preferred features and visualization methods \citep{zeiler_visualizing_2013,simonyan_deep_2014}. These approaches treat features as encoding their causal inputs, whereas we shift focus to the information that can be read out from features irrespective of origin.

\textbf{Recoverability-based approaches.} A complementary line of work investigates representation by what can be recovered. Feature inversion demonstrates that fine-grained input details persist in higher network layers \citep{mahendran_understanding_2014}, and probing uses linear classifiers to quantify decodable content \citep{alain_understanding_2018}.
Brain decoding and reconstruction studies have shown that various perceptual experiences can be readout from neural activity \citep{kamitani_decoding_2005,miyawaki2008visual, horikawa_generic_2017,shen_deep_2019}, including those decoupled from causal origin such as attended stimuli \citep{kamitani_decoding_2005,horikawa_attention_2022}, visual illusion \citep{cheng_reconstructing_2023}, mental imagery \citep{shen_deep_2019}, and dream \citep{horikawa2013neural,horikawa_hierarchical_2017}.
These studies reveal the decodable content from neural activity, whereas our research focuses on formalizing representations from an informational viewpoint by utilizing these recovery techniques as a computational procedure. We offer the quantification of representation size at the instance level through the formalization of readout representations.

\textbf{Redefining neural representations.}
Studies have proposed frameworks for redefining neural representations by the information that drives behavior \citep{panzeri2017cracking}, or by the informational properties \citep{pohl2025clarifying}. \citet{panzeri2017cracking} is specifically focused on behavior, while ours incorporate arbitrary readout procedures, not limited to behavior. Unlike these studies, our framework offers a coherent handling of representations decoupled from causal origin, such as misrepresentation and dreaming, and offers a quantitative metric to characterize the robustness of representations.

\textbf{Redundancy and robustness in neural representations.} Studies have proposed metrics to characterize neural representation through redundancy using mutual information and partial information decomposition (PID) \citep{schneidman2003synergy,williams2010nonnegative}, and through compression \citep{tishby2015deep,shwartz2017opening}. These approaches typically analyze redundancy and compression at the dataset level, whereas our representation size metric quantifies redundancy at the instance level, offering new insights into how individual inputs are robustly encoded.
In terms of robustness, adversarial examples demonstrates that small perturbations can drastically alter encodings \citep{szegedy_intriguing_2014}, and are attributed to the presence of features rather than artifacts \citep{ilyas_adversarial_2019}.
Probabilistic generative models, represented by VAE \citep{kingma2013auto,rezende2014stochastic}, explicitly introduce variability into representations in artificial neural networks.
While these approaches offer insights on how to incorporate variability into representations, our work focuses on analyzing the extent of variability inherently present even in deterministic models trained for standard tasks.

\section{Readout representation}
We introduce the readout representation framework, which formalizes the informational view of representation. Consider the brain or neural network \(f: X \times \Xi \rightarrow H\) that maps an input stimulus \(x \in X\) to neural features \(h \in H\) under the brain state or context \(\xi \in \Xi\).
Let \(S\) denote a signal space of interest, and let the true signal corresponding to the input \(x\) be given by a reference mapping \(\bar\pi: X \rightarrow S\). The signal space \(S\) may represent the input itself, a compressed description, or any latent variables of interest. Let \(\pi: H \rightarrow S\) denote a readout procedure, which extracts signal information from features.

Under the causal view, a feature \(h \in H\) is considered a representation of the input \(x \in X\) that causally generated it, as \(h = f(x)\). When there is randomness between \(x\) and \(h\), such as in brain activity or VAE, one typically considers the expectation \(\mathbb{E}_{\xi}[f(x, \xi)]\). In contrast, the information view situates the feature \(h\) as a representation of \(s\) based on the information it conveys.

\subsection{Definition of readout representation}
First, following the informational view, representation is defined by the information content recoverable from the feature using a readout method \(\pi\), independently of its causal origin.

\begin{definition}[Representation]
    \(h \in H \text{ represents }  s \in S \Longleftrightarrow s = \pi(h).\)
\end{definition}

\noindent This definition contrasts with the approach commonly considered in the hierarchical-causal view, which defines the feature \(h\) as a representation of its causal origin \(x\).
Under this definition, multiple features may represent the same signal, which we term \textit{readout representations}.

\begin{definition}[Readout representation]
    The set \(H_{s}^{\pi} \coloneqq \{ h \in H \mid \pi(h) = s \}\) denotes readout representations of a signal \(s\).
\end{definition}

\noindent Given the set \(H_s^\pi\), we can evaluate the spread of the set \(H_s^{\pi}\) as \textit{representation size}, which can be interpreted as the volume occupied by specific information within the feature space. The representation size can be instantiated in various ways, such as cardinality, geometric volume, or other distance-based or probabilistic indicators.

Note that the readout representation is a natural extension of the causal representation in the specific case where the readout is the inverse of the encoding function.
\begin{remark}
    Suppose \(S \coloneqq X\) and \(f\) is injective, and ignore the context \(\xi\). Define \(\pi(h) \coloneqq f^{-1}(h)\) for \(h \in \operatorname{Im} f\), and define \(\pi(h)\) to be an arbitrary element of \(X\) otherwise. Then the causal representation \(h = f(x)\) lies in the readout representation \(H_{x}^{\pi}\).
    Even in this case, any other feature \(h^\prime \neq h\) is also included in \(H_{x}^{\pi}\) as long as \(\pi(h^\prime) = \pi(h)\).
\end{remark}

Readout representation enables us to handle \emph{misrepresentation} as a situation where the extracted signal differs from the corresponding true signal. To ensure that the feature space represents nontrivial signals, we first introduce \textit{representation capability} as an ability of \((f, \pi)\) to distinguish signals in order to exclude trivial cases where all features represent the same signal, regardless of the input.

\begin{definition}[Representation capability]
    \((f, \pi)\) has representation capability of \(S\) if and only if
    \[
    \exists x_1, x_2 \in X, \exists \xi \in \Xi \quad \text{s.t.} \quad \bar\pi(x_1) \neq \bar\pi(x_2), \bar\pi(x_1) = \pi \circ f(x_1, \xi), \bar\pi(x_2)=\pi \circ f (x_2, \xi).
    \]
\end{definition}
\begin{definition}[Misrepresentation]
Suppose \((f, \pi)\) has representation capability of \(S\), and feature \(h = f(x)\) represents \(s = \pi(h)\). Then, we say \(h \text{ misrepresents } s \text{ if and only if } \pi(h) \neq \bar \pi (x)\).
\end{definition}

\subsection{Case studies}
Here, we demonstrate in the following case studies how readout representation accommodates situations that challenge the causal view, including misrepresentation and dreaming.

\textbf{Misclassification.}
Suppose a subject sees a rope but mistakenly reports a snake. Let the input space \(X\) be natural scenes, the feature space \(H\) be neural activity in the visual cortex, and the signal space \(S\) be object categories of attended objects. The reference mapping \(\bar\pi\) assigns the true category, and the readout procedure \(\pi\) corresponds to the subject's report. In causal view, the feature \(h = f(x, \xi)\) (or \(\mathbb{E}[f(x, \xi)]\)) should be treated as a representation of the rope, independent of the subject's report, and thus it cannot account for misrepresentation. However, in our framework, we can say that the feature \(h = f(x, \xi)\) misrepresents the rope as a snake when \(\pi(h) = \text{snake} \neq \text{rope}\). The same framework also covers a case where the subject views the scene again under a different state \(\xi'\) and reports ``rope,'' showing that both correct and incorrect outcomes are naturally part of the representation.

\textbf{Illusion.}
Consider a Müller-Lyer figure, where two parallel lines of equal length appear different. Let \(X\) be images of two parallel lines with length difference \(d(x)\), and let \(S\) indicate whether this difference exceeds a perceptual threshold \(\Delta\). The reference mapping is \(\bar\pi(x) = \mathbbm{1}_{d(x)>\Delta}\), while the readout \(\pi\) is the subject’s report. For ordinary images, the subject correctly classifies differences, so representation capability is satisfied. For the illusion image \(x\), we have \(\bar\pi(x)=0\) (no actual difference), but the subject may report \(\pi \circ f(x,\xi)=1\). Within our framework, this simply means that the feature \(h=f(x,\xi)\) represents the signal ``longer line'' despite the absence of a physical difference. The same system correctly represents non-illusory cases, and illusory responses appear as part of the same representational structure rather than as contradictions.

\textbf{Dreaming.}
During sleep, neural activity can generate representations without external input. Let \(X\) be visual stimuli and \(S \coloneqq X\). Under wakefulness (\(\xi_w\)), the subject can describe stimuli through the readout \(\pi_o\), establishing representation capability. Under sleep (\(\xi_s\)), there is no external input, so \(\bar\pi(x)=\text{no stimulus}\). If the subject dreams of a dog, then \(\pi_o \circ f(x,\xi_s)=\text{dog}\). Here the feature \(h=f(x,\xi_s)\) represents content decoupled from the environment, not a misrepresentation in the strict sense. Similarly, consider a brain decoder \(\pi_d\) trained on wakeful activity as in \citep{horikawa_hierarchical_2017}. The decoder may output a dog when applied to the sleep activity, again producing a valid representation even in the absence of external inputs. The framework thus treats dream content as part of the representational space defined by readout, integrating it into the same formal structure that applies to waking perception.

\section{Experiments}
We examine the informational content of deep neural networks using readout representation. Specifically, we demonstrate that, contrary to the idea that information is progressively discarded, crucial input details are robustly preserved across a broad range of hierarchical features. First, we show that information about an input stimulus is often represented in a broad area of the feature space in multiple models, modalities, and layers. We show this by demonstrating that the input information is recoverable from heavily perturbed features. Second, we show the potential of the representation size metric to highlight the representational properties of different input stimuli, model architecture, and model size. Finally, we experimented with a simplified model to highlight the role of redundancy in the robust recovery from perturbed features.

The following experimental settings and methodologies were applied across experiments unless otherwise specified. We provide details in \Cref{sec:experiment}.

\textbf{Feature inversion.} We instantiate the readout \(\pi\) with feature inversion because it makes minimal structural assumptions about \(\pi\) and searches the preimage of \(f\) directly: if input-level information is present anywhere in the feature, inversion will retrieve it without relying on task-specific decoders. Concretely, given a target feature \(h\), we define \(\pi(h) = \arg\min_{x} L\bigl(f(x), h\bigr)\) where \(L\) is a feature-matching loss. For vision models, we optionally regularize inversion with Deep Image Prior (DIP) \citep{ulyanov_deep_2020} to reduce high-frequency artifacts while avoiding external pretrained generative priors. We also show the robustness of our findings without DIP in \Cref{sec:ablation_dip}. For language models, we optimize token logits, which are converted to embeddings and fed into the model, iteratively minimizing the distance to the target features.

\textbf{Distance and accuracy measures.} We use correlation distance\footnote{Although correlation distance (defined as \(1 - \text{correlation coefficient}\)) does not lead to a proper distance metric, it suffices for our purposes of ranking similarity consistently across heterogeneous spaces.} as a distance function in the feature space \(d_H\) across both modalities, and as a distance function in the input space \(d_X\) for the vision modality because it yields a unit-free, dimension-agnostic scale that enables cross-layer/modality comparisons. We verify robustness to the choice of distance by repeating evaluations with perceptual pixel-space metrics (SSIM \citep{zhou_image_2004}, PSNR, LPIPS \citep{zhang2018perceptual}, and DISTS \citep{ding_image_2022}) and observing the same qualitative trends (\Cref{sec:results_vision}). For the language modality, we use token error rate as a distance measure in input space, defined as the proportion of tokens in the reconstructed text that differ from the original text.

\textbf{Readout representation.} Let \(d_X\) and \(d_H\) denote distance functions in the input and feature spaces, respectively. To account for minor deviations due to numerical inaccuracies in the optimization process, we use a relaxed version of readout representations parameterized by a threshold \(t \geq 0\) on input-space distances: \(H_{x, t}^{\pi} = \{ h \in H \mid \forall x' \in \pi(h), d_X(x, x') < t \}\). The threshold \(t\) is chosen as a sufficiently small value specific to the modality and distance measure: \(0.1\) correlation distance in vision, and \(0.3\) token error rate in the language modality. We instantiate the representation size by the maximum feature-space deviation within the relaxed set: \(r_x = \max \{ d_H(h, f(x)) \mid h \in H_{x,t}^{\pi} \}\).

\textbf{Feature perturbation.} As a feature perturbation, we added Gaussian noise to the original feature across ten noise levels, calibrated to produce specific correlation distances from the original feature.

\begin{figure}[t]
  \centering
  \phantomsection         
  \includegraphics[width=1.0\linewidth]{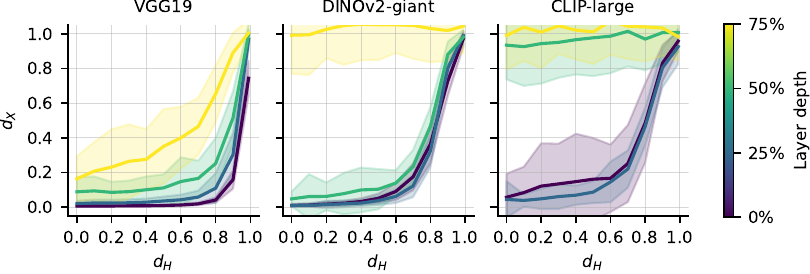}
  \caption{
      Images can be accurately recovered from heavily perturbed features. The x-axis shows the correlation distance between the original and perturbed features, and the y-axis shows the correlation distance between the original and reconstructed images. Each line represents a quarter-depth layer, colored by layer depth. Shaded areas indicate $\pm1$ SD across images: for each distance bin and image, we computed the standard deviation across images. \textbf{Left}: In VGG19, lower to middle layers retained high-fidelity information (within 0.1 correlation distance) even after 0.7 correlation distance perturbation. \textbf{Middle} and \textbf{Right}: Similar patterns were observed in lower to middle layers of DINOv2 and lower layers of CLIP. Full results are provided in the \Cref{sec:results_vision}.
      }
    \label{fig:fig2_vision}
\end{figure}

\subsection{Feature inversion from perturbed features}
We first examined how extensively an input is represented in the feature space of neural models by reconstructing inputs from perturbed features. We conducted experiments across both vision and language modalities. For vision, we mainly used VGG19 \citep{simonyan_very_2015}, CLIP \citep{radford_learning_2021}, and DINOv2 \citep{oquab_dinov2_2024}, spanning both convolutional neural networks (CNNs) and Vision Transformers (ViTs). We also performed experiments using a variational autoencoder and report results in \Cref{sec:results_vision}. For language, we used the BERT \citep{devlin_bert_2019}, GPT2-series \citep{radford_language_nodate}, and OPT-series \citep{zhang_opt_2022}. In both modalities, we performed reconstruction of 64 samples, randomly sampled images from ImageNet \citep{deng_imagenet:_2009} and texts from C4 \citep{Raffel_explore_2019}. We note that our aim is to illustrate how multiple features can represent a single input, rather than to benchmark model performance. Features were extracted from multiple intermediate layers for each model---all convolutional layers in VGG19, and every quarter-depth transformer block in ViT-based models and language models.

We found that the original images and text could be reliably reconstructed even from significantly perturbed features in lower to middle layers in multiple models (Figure \hyperref[fig:fig1_concept]{1B}; see \Cref{sec:results_vision,sec:results_language} for other images/texts). For example, in the lower to middle convolutional layers of VGG19, near-perfect recovery was observed at feature correlation distances up to 0.8. Quantitatively, in VGG19 lower layers, features perturbed up to 0.7 correlation distance still yielded reconstructions within 0.1 correlation distance in pixel space (\Cref{fig:fig2_vision}). Similarly, the high fidelity recovery from perturbed features was also observed in the lower to middle layer feature in DINOv2 and CLIP models. Full results, including reconstructed images from all models, can be found in \Cref{sec:results_vision}.

\begin{figure}[t]
  \centering
  \phantomsection
  \includegraphics[width=1.0\linewidth]{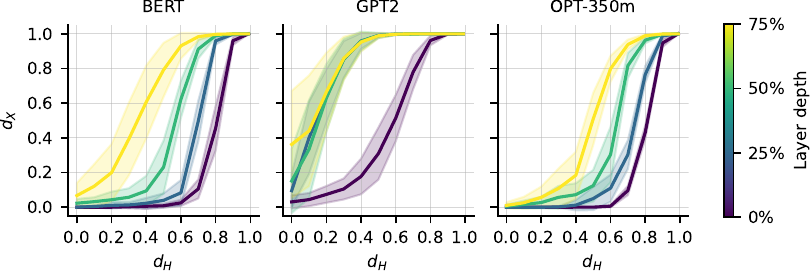}
  \caption{
      Input texts can often be recovered with high accuracy from perturbed features. The x-axis shows the correlation distance between the original and perturbed features, and the y-axis shows the token error rate. Each line shows results for features from a quarter-depth layer, with its color showing the layer depth. Shaded regions indicate $\pm1$ SD across samples. \textbf{Left}: Similar to the vision modality, the input tokens are recovered with high accuracy even from heavily perturbed lower to middle layer features. \textbf{Middle}: In GPT2, we do not observe high fidelity recovery from perturbed features, although the accuracy was substantially above chance levels across most feature distances. \textbf{Right}: Lower to middle OPT-350m layers show extended readout representations, as well as BERT. We provide further results in \Cref{sec:results_language}.
      }
    \label{fig:fig3_language}
\end{figure}

A similar pattern was observed in some of the language models (\Cref{fig:fig3_language}). Input tokens were recoverable significantly above chance from perturbed features across models. In particular, BERT and OPT-350m showed near-perfect recovery even at high perturbation levels (up to 0.7 correlation distance) in lower layers. GPT2 and some OPT variants exhibited reduced performance, but still exceeded chance to certain perturbations. Please note that, given the large vocabulary sizes (BERT: 30,522; GPT2: 50,257; OPT-350m: 50,272), random guessing yields near-zero accuracy. Additional results, including comparisons across model sizes, recovered sentences, and analyses of model differences, are provided in \Cref{sec:results_language}.

Together, these results illustrate that the representation of an identical or nearly identical stimuli often extends to a broad region in the feature space, a property observed across multiple architectures, modalities, and layers. Such extensive representational coverage suggests inherent redundancy in neural representations, enabling accurate information recovery even from perturbed features.

\subsection{Application of representation size to instance-level analysis}
Next, we use representation size to clarify how input stimuli are represented at the instance level in the hierarchical features, using VGG19 as a case study. Specifically, we compared representation sizes across correctly and incorrectly classified images, and between natural and noise images.

\begin{figure}[t]
  \centering
  \phantomsection
  \includegraphics[width=1.0\linewidth]{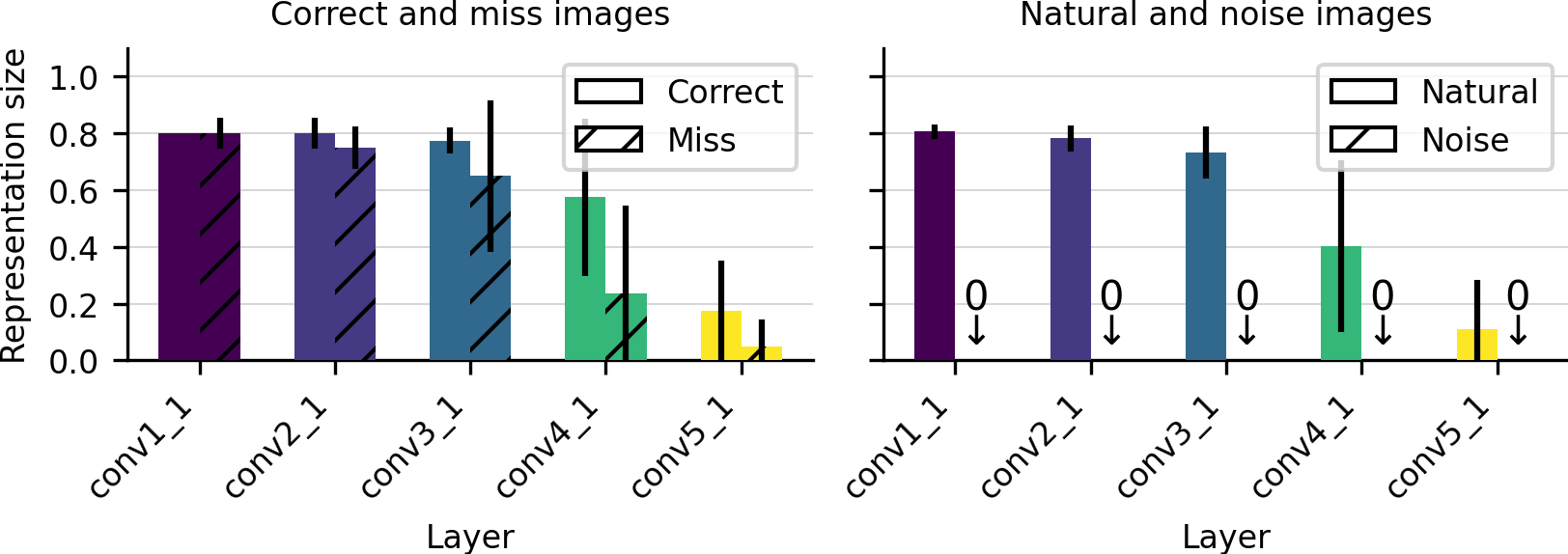}
  \caption{
    Application of representation size. Representation sizes across VGG layers for different input types. Error bars show $\pm1$ SD across samples. \textbf{Left}: Correctly classified images exhibit larger representation sizes than misclassified ones, particularly in deeper layers. This suggests that successful classification is associated with more redundant and robust representations. \textbf{Right}: Comparison between natural and noise images. Noise images show zero representation sizes. 
    }
    \label{fig:fig4_image_types}
\end{figure}

First, we evaluated the representation size for images correctly and incorrectly classified by VGG19 on the ImageNet \citep{deng_imagenet:_2009} validation set. We sampled 8 images each from correctly classified and misclassified examples. Correct classifications were defined by correct top-1 prediction, and misclassifications by failure to include the correct label in the top-5 predictions. Results showed that correctly classified images exhibited consistently larger representation sizes, particularly in later layers (\Cref{fig:fig4_image_types} left), indicating the connection of the representation size to the model performance.

Second, we compared the representation size between natural images and noise images. Noise images were generated by sampling pixel values from a uniform distribution. Noise images exhibited zero representation sizes across all layers, in contrast to the broader representations seen for natural inputs (\Cref{fig:fig4_image_types} right).
The above results suggest that the size that a particular instance occupies in the model's internal representation is related to the nature of that instance and whether the model was able to learn it effectively, and reflects the model's perceptual reliability.
This highlights the potential effectiveness of the metric for diagnosing model inference results for specific instances.
Note that this analysis differs fundamentally from quantifying the amount of information in causal representations using such as mutual information, as it enables the calculation of the size of a single instance representation even for neural networks exhibiting deterministic behavior.

\subsection{Interpretation of readout representation}
What factors underlie the extended readout representations?
First, we analyze how the representation size relates to the feature dimensionality across models and layers. Second, we use a simplified toy model to illustrate how high-dimensional mappings naturally yield redundant, robust representations that support stable readout even under perturbation.

\begin{figure}[t]
  \centering
  \phantomsection
  \includegraphics[width=1.0\linewidth]{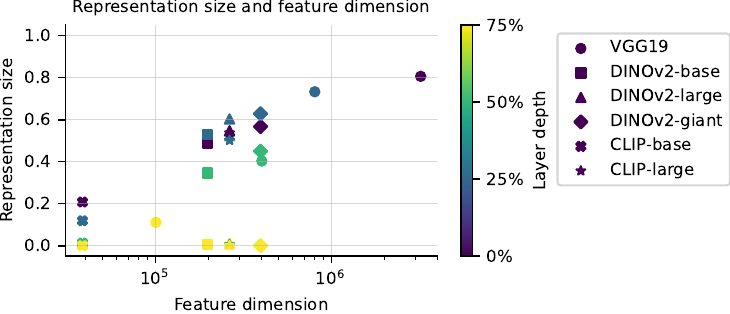}
  \caption{
    Relationship between representation size and feature dimension across models and layers. Each point represents a layer from each model, plotted by its feature dimensionality (x-axis, log scale) and corresponding representation size (y-axis). Color indicates relative layer depth, and marker shape denotes model identity. Mid-depth layers with moderate feature dimensions tend to exhibit larger representation sizes, suggesting a trade-off between representational richness and compressibility across network hierarchies.
    }
    \label{fig:fig5_dim_and_size}
\end{figure}

First, we examined the relationship between feature dimensionality and representation size across layers of various vision models (\Cref{fig:fig5_dim_and_size} right). Overall, we observed that layers with higher feature dimensionality tended to exhibit larger representation sizes. However, this trend weakened in the higher layers. These results suggest that redundancy enabled by high-dimensional representations is a key factor in representation size, while deeper layers may trade off this redundancy for compactness and task-specific abstraction.
The size of representations in higher layers may be influenced by the characteristics of the labeled data and objective function used in model training.
Consistent with this, when we ablate the network and assess representation size directly in pixel space, the size is substantially smaller than in mid-VGG layers (\Cref{sec:ablation_dip}), indicating that the observed high representation size cannot be attributed to the DIP alone. This motivates the toy model analysis, which isolates the role of redundancy under controlled conditions.

\begin{figure}[t]
  \centering
  \phantomsection
  \includegraphics[width=1\linewidth]{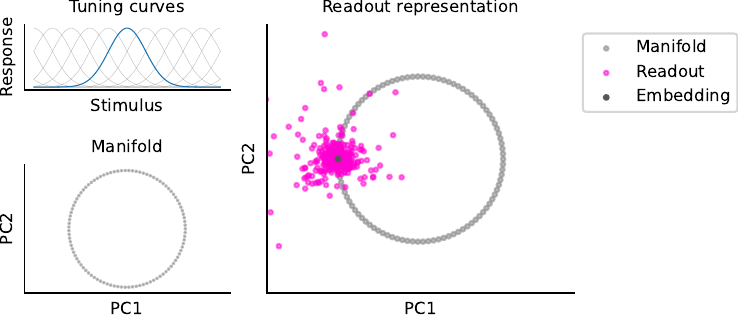}
  \caption{
    Readout representations in a toy neural system with bell-shaped tuning curves. \textbf{Top left}: Tuning curves of neurons. 10 out of 100 neurons are shown. Each neuron selectively responds to different locations in a one-dimensional stimulus space. \textbf{Bottom left}: The resulting neural manifold, constructed by projecting the neural responses of all stimuli onto the first two principal components. \textbf{Right}: The readout representations of a single input (a dark gray dot) visualized by magenta points. Each readout representation point shows a perturbed feature from which the original input was successfully readout, showing that multiple features represent the same signal.
  }
  \label{fig:fig6_toy_model}
\end{figure}

Second, we examined the readout representation of a simplified toy model to illustrate the emergence of extended representations. The model consisted of 100 neurons with bell-shaped tuning curves distributed evenly across a one-dimensional input space $X = [0,1)$. Each stimulus $x \in X$ was projected onto a higher-dimensional neural feature space $h = f(x)$ (\Cref{fig:fig6_toy_model}, top left).
Due to the higher dimensionality of the feature space compared to the input, there is redundancy in representation, and the inputs were embedded into a manifold with low intrinsic dimension (\Cref{fig:fig6_toy_model}, bottom left). This redundancy would allow robust recovery: even if a feature vector deviates from the original manifold, the original input would still be recovered.
To test this expectation, we perturbed the feature vectors by adding noise and attempted to recover the original inputs by retrieving the closest features to the perturbed points. On the principal component (PC) space, we projected the points from which we could readout the original input (\Cref{fig:fig6_toy_model}, right). Projected points are distributed in a broad region of the PC space, showing that multiple features represent the same signal. Although this toy model is simplified, a similar mechanism is likely at work in deep neural networks, given evidence of low intrinsic dimensionality relative to their feature spaces \citep{pope_intrinsic_2021}.

\section{Discussion}
Our study introduced readout representation, a framework that defines representations by the information recoverable from features rather than their causal origin. This perspective helps resolve the tension between hierarchical abstraction and detail retention: while abstraction emphasizes categorical or task-relevant features, our results show that fine-grained input information remains recoverable across broad regions of feature space. Through experiments in both vision and language models, we demonstrated that inputs can be reconstructed from heavily perturbed features, indicating that representations are not single points but extended sets. To quantify this property, we proposed representation size, which measures the breadth of features that support accurate recovery and links redundancy to robustness and performance. Together, these findings suggest that neural representations are inherently information-rich and resilient, challenging the assumption that abstraction necessarily discards input details.
The toy system concretely demonstrates how high-dimensional embeddings produce extended readout sets, providing a mechanistic explanation for the empirical trends.

In relation to our original motivation: (i) we quantified recoverability via representation size, (ii) we linked it to the performance and redundancy, and (iii) a simplified model accounted for the possible mechanism on how representations can retain recoverable detail while supporting hierarchical abstraction.
This work offers several future directions. First, the representation size provides a method for analyzing neural representations. The representation size offers a per-sample analysis of representational redundancy and robustness, complementing existing methods that typically analyze representations at the dataset level \citep{schneidman2003synergy, denil_predicting_2013, feather_model_2023, kriegeskorte_representational_2008,kornblith_similarity_2019,huh_platonic_2024}. An additional direction is to apply the readout framework to biological data, extending current decoding and reconstruction studies. Brain decoding studies often predict deep network features from noisy brain activity \citep{shen_deep_2019,cheng_reconstructing_2023}, and their success may be explained by the large representation size of the networks employed. Building models with even larger representation sizes could potentially improve decoding accuracy.

\section*{Acknowledgments}
We thank our laboratory members for their valuable feedback. This work was supported by Japan Society for the Promotion of Science (JSPS: KAKENHI grants JP20H05954, JP20H05705, JP25H00450 to Y.K., and JP21K17821 to Y.N.), Japan Science and Technology Agency (JST: CREST grants JPMJCR22P3 to Y.K.), New Energy and Industrial Technology Development Organization (NEDO: commissioned project, JPNP20006 to Y.K.), and Guardian Robot Project, RIKEN.

\newpage
\begin{appendices}

\section{Additional information}
We provide additional information about our study. \Cref{sec:limitations} describes the limitations of our work. \Cref{sec:experiment} describes the details of the experiment including procedures for feature inversion, feature perturbation, optimization settings, and model specifications. \Cref{sec:results_vision} provides additional results for vision models, including reconstructed images and quantitative results. \Cref{sec:results_language} reports the additional results for language models, including reconstructed sentences, quantitative evaluations, and insights into the difference between models.

\section{Limitations}
\label{sec:limitations}
Despite these contributions, several limitations remain. First, the source of variation in representation size across architectures remains unclear. For example, language models such as GPT2-Large and OPT-1.3b showed lower recovery performance under perturbation compared to others (\Cref{sec:results_language}). In vision models, we found that feature dimensionality has a strong effect on representation size, and we also provide additional analysis of performance differences among language models in the \Cref{subsec:lm_rep_size}, though a comprehensive understanding remains an open question. Second, while we propose representation size as a metric for model evaluation, our experiments demonstrate only preliminary use cases; its broader utility remains to be established. Third, although our framework is inspired by both artificial and biological systems, its empirical relevance to neuroscience remains untested. Finally, our implementation adopts feature inversion as the readout method, which introduces some computational overhead. However, our proposed framework is independent from the choice of readout method, and users may adopt alternative procedures better suited to their needs and computational budgets.

\section{Experiment details}
\label{sec:experiment}
\subsection{Feature inversion}
\subsubsection{Problem formulation}
\label{sec:problem_formulation}
In our feature inversion experiments, we aim to recover the original input from a given neural representation, referred to as the target feature. Feature inversion reverses the encoding process by identifying an input that produces a feature vector similar to the target. Let \( f: X \rightarrow H \) denote a neural network that maps an input \( x \in X \) to its corresponding feature \( h \in H \). Given a target feature \( h \in H \simeq \mathbb{R}^d \), the feature inversion is formulated as finding the set of inputs whose features minimize a predefined loss:
\[
\pi(h) \coloneqq \arg\min_{x \in X} \mathcal{L}(f(x), h),
\]
where \( \mathcal{L} \) denotes a loss function that quantifies the dissimilarity between the feature vector \( f(x) \) and the target feature \( h \). The solution set \( \pi(h) \) contains inputs that yield features close to the target feature.

For the loss function, we used the mean squared error (MSE) in vision models:
\[
\mathcal{L}(h_{\text{recon}}, h_{\text{target}}) = \|h_{\text{target}} - h_{\text{recon}}\|_2^2,
\]
and the linear combination of MSE and cosine loss in language models:
\[
\mathcal{L}(h_{\text{recon}}, h_{\text{target}}) = \|h_{\text{target}} - h_{\text{recon}}\|_2^2 - \cos(h_{\text{target}}, h_{\text{recon}}).
\]

\subsubsection{Methods}
To solve the minimization problem defined in \Cref{sec:problem_formulation}, we adopt gradient descent optimization. Instead of directly optimizing the input \(x \in X\), we  optimize it through iteratively optimizing a latent variable \(z \in Z\). Specifically, we define a generator function \(g: Z \rightarrow X\) that maps the latent variable to the input space. Each step of the optimization process is as follows:
\[
z \leftarrow z - \eta \nabla_z \mathcal{L}(f(g(z)), h), \quad x \leftarrow g(z),
\]
where \(\eta\) is the learning rate.

\textbf{Vision models.} We use Deep Image Prior (DIP) \citep{ulyanov_deep_2020} as the generator \(g\). Its parameter \(z\) is optimized to produce the input image \(x\) from a fixed random inputs. The architectural bias of DIP acts as a structural prior, suppressing high-frequency artifacts and improving perceptual quality. We choose DIP because it poses minimal prior assumptions about the image distribution and does not require any training data, making it suitable for our analysis. Specifically, we deliberately avoided using pre-trained generative models as a prior, such as GANs and diffusion models, which may introduce biases from their training data and limit the generality of our findings. As an ablation, we also performed feature inversion without DIP and report results in the \Cref{sec:ablation_dip}.

\textbf{Language models.} For language models, the input is a discrete token sequence \(x \in X = \{1, \ldots, V\}^T\), where \(V\) denotes the vocabulary size and \(T\) the sequence length. For the readout operation, we relax this discrete space to continuous space since the gradient-based optimization in discrete space is not trivial.
Instead of directly optimizing the input tokens, we optimize the token logits \(z \in \mathbb{R}^{T \times V}\), which are then converted to the continuous tokens \(x \in \mathbb{R}^{T \times V}\) as
\[
    x_i = g(z_i) = \operatorname{softmax}(z_i), \quad \text{for } i = 1, \dots, T,
\]
where the generator function \(g\) applies a row-wise softmax to \(z\) and each \(x_i\) represents the relaxed categorical distribution over the vocabulary at position \(i \in [T]\).
The final token sequence after gradient-based optimization is obtained by taking the argmax over each row of the optimized logit matrix:
\[
x_i = \arg\max_j z_{i,j}, \quad \text{for } i = 1, \dots, T.
\]

\subsubsection{Optimization settings}
In both vision and language modalities, we used the AdamW optimizer from PyTorch (version 2.3.1). \Cref{tab:opt_params} summarizes the optimizer settings and training configurations used for feature inversion across modalities. Default PyTorch parameters were used for AdamW if not specified. In vision modality, we used a linear learning rate scheduler, which decayed the learning rate to zero over the course of training.

\begin{table}[h]
    \caption{Optimizer and training parameters used in feature inversion experiments.}
    \centering
    \begin{tabular}{lccc}
        \toprule
        \textbf{Parameter} & \ \textbf{Vision} & \textbf{Language} \\
        \midrule
        Optimization Target & DIP Latent & Token Logits \\
        Learning Rate & 0.0001 & 0.1 \\
        Iterations & 10,000 & 10,000 \\
        \bottomrule
    \end{tabular}
    \label{tab:opt_params}
\end{table}

\subsubsection{Feature perturbation}
To systematically evaluate the extent of representations, we perturbed feature vectors by adding Gaussian noise with calibrated variance. Given a feature vector $h \in \mathbb{R}^d$, we generated a perturbed feature $h' = h + \varepsilon$, where $\varepsilon \sim \mathcal{N}(0, \sigma^2 I_d)$.

The variance $\sigma^2$ was analytically determined to produce a target correlation distance $d_H(h, h^\prime) = c \in (0, 1)$ between the original and perturbed features. Under high-dimensional assumptions, the expected correlation distance can be approximated as:
\[
c \approx 1 - \frac{1}{\sqrt{1 + \frac{\sigma^2}{\operatorname{Var}(h)}}},
\]
where, \(\operatorname{Var}(h)\) denotes sample variance of the feature vector \(h\). Solving for $\sigma^2$ gives:
\[
\sigma^2 \approx \operatorname{Var}(h) \left( \frac{1}{(1 - c)^2} - 1 \right).
\]
This allowed control over perturbation magnitude. We used 10 correlation distance levels: $\{0.1, 0.2, 0.3, 0.4, 0.5, 0.6, 0.7, 0.8, 0.9, 0.99\}$.

\subsubsection{Computational resources}
Our feature inversion experiments flexibly scale with available resources by adjusting key parameters such as the number of noise levels, random seeds, and the number of parallel execution. As a reference, a minimal experiment—reconstructing a single image from one VGG19 layer with 10 noise levels, a number of parallel execution of 10, and one random seed—requires approximately 10 GB of GPU memory and 20 minutes of runtime on an NVIDIA V100 GPU.

Full-scale experiments, including all layers, models, and repetitions, were conducted using the following hardware:
\begin{itemize}
    \item \textbf{Local resources:} NVIDIA Tesla V100S (32 GB), Quadro RTX 8000 (48 GB), RTX A6000 (48 GB), and A100 GPUs.
    \item \textbf{Cloud resources:} AWS g5.48xlarge instances equipped with 8 NVIDIA A10G Tensor Core GPUs, ABCI (AI Bridging Cloud Infrastructure) rt\_HF nodes equipped with 8 NVIDIA H200 SXM 141GB GPUs.
\end{itemize}

\subsection{Feature forwarding}
In our supplementary experiment using a variational autoencoder (VAE) model, we additionally adopted another readout method which we call feature forwarding. Given the neural features of the VAE, we directly pass it through the decoder module of the model to reconstruct the image from it. This approach leverages the learned generative capabilities of the VAE, providing an alternative to optimization-based feature inversion. We provide results using this method in \Cref{sec:results_vision}.

\subsection{Model specifications}
We evaluated feature representations across a range of pretrained vision and language models.
For vision modalities, we used VGG19, CLIP (Base Patch32, Large Patch14), DINOv2 (Base, Large, Giant), and SDXL-VAE. For language modalities, we used BERT (Base), GPT2 (Small, Medium, Large, XL), and OPT (125M, 350M, 1.3B, 2.7B). Except for VGG19, all models were obtained from the HuggingFace Transformers library using publicly available checkpoints. For each model, we selected representative layers for analysis: all 16 convolutional layers in VGG19, and one layer per quarter depth in other models (e.g., layers 0, 3, 6, 9 in a 12-layer model). Below, we provide details on the source of each model.

\textbf{VGG19.} We used the original Caffe weights from \url{http://www.robots.ox.ac.uk/~vgg/software/very_deep/caffe/VGG_ILSVRC_19_layers.caffemodel}, which we converted to PyTorch format. The converted weights are available at \url{https://figshare.com/ndownloader/files/38225868}.

\textbf{Other models.} \Cref{tab:hf_id} lists HuggingFace model identifiers for each model.

\begin{table}[h]
    \caption{HuggingFace model identifiers for the transformer models used in our experiments.}
    \centering
    \begin{tabular}{ll}
        \toprule
        \textbf{Model Family} & \textbf{HuggingFace Identifier} \\
        \midrule
        DINOv2 (ViT) & facebook/dinov2-base \\
        & facebook/dinov2-large \\
        & facebook/dinov2-giant \\
        CLIP (ViT)   & openai/clip-vit-base-patch32 \\
        & openai/clip-vit-large-patch14 \\
        SDXL-VAE (VAE) & stabilityai/sdxl-vae \\
        BERT         & google-bert/bert-base-uncased \\
        GPT2         & openai-community/gpt2 \\
        & openai-community/gpt2-medium \\
        & openai-community/gpt2-large \\
        & openai-community/gpt2-xl \\
        OPT          & facebook/opt-125m \\
        & facebook/opt-350m \\
        & facebook/opt-1.3b \\
        & facebook/opt-2.7b \\
        \bottomrule
    \end{tabular}
    \label{tab:hf_id}
\end{table}

\subsection{Dataset}
\begin{figure}[H]
    \centering
    \includegraphics[width=1.0\linewidth]{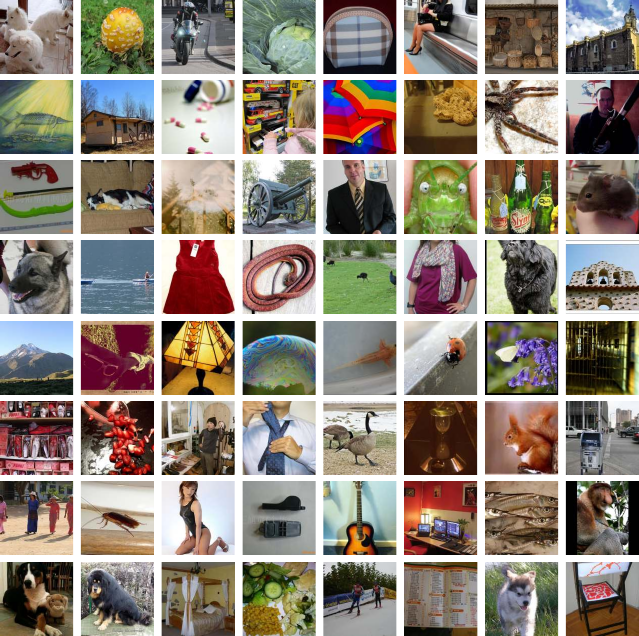}
    \caption{
        64 natural images used in the vision experiments, randomly sampled from the test-split of ImageNet dataset. The images were selected prior to analysis and used consistently across experiments without post-hoc cherry-picking.
    }
    \label{fig:natural_images}
\end{figure}

For the vision modality, we randomly sampled 64 natural images for our primary experiments (\Cref{fig:natural_images}) from the test-split of the ImageNet dataset via HuggingFace datasets (ILSVRC/imagenet-1k).

\begin{figure}[H]
    \centering
    \includegraphics[width=1.0\linewidth]{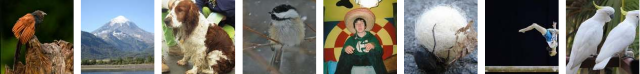}
    \caption{Images that are correctly classified by VGG19 on the ImageNet validation set. Each image was correctly classified based on the top-1 prediction of VGG19.}
    \label{fig:correct}
\end{figure}

\begin{figure}[H]
    \centering
    \includegraphics[width=1.0\linewidth]{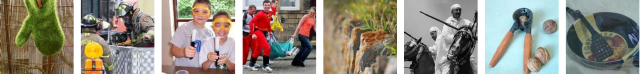}
    \caption{Images that are incorrectly classified by VGG19 on the ImageNet validation set. Each image was misclassified as the correct label was not included in the top-5 predictions of VGG19.}
    \label{fig:incorrect}
\end{figure}

To analyze differences in representation size between correctly classified images (\Cref{fig:correct}) and incorrectly classified ones (\Cref{fig:incorrect}), we sampled 16 images from the ImageNet validation set in total. Correctly classified examples included eight images where the model's top-1 prediction matched the ground-truth label. Misclassified examples included eight images where the true label was not within the top-5 predictions. We excluded ambiguous cases to ensure label clarity.

\begin{figure}[H]
    \centering
    \includegraphics[width=1.0\linewidth]{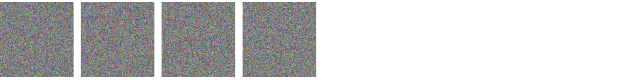}
    \caption{Random noise images used for representation size analysis. Each image was generated by independently sampling pixel values from a uniform distribution over [0, 255] with different random seeds.}
    \label{fig:noise_images}
\end{figure}

For the experiment with noise images, we generated noise images by sampling pixel values uniformly from [0, 255] (\Cref{fig:noise_images}). We prepared four images with different random seeds to improve robustness.

For the language modality, we sampled 64 sequences from the validation split of the C4 dataset via HuggingFace Datasets (allenai/c4). Each sequence was truncated to a maximum of 256 tokens in order to control computational costs.

In all experiments, the dataset was curated before analysis, and no post-hoc selection was performed to ensure unbiased evaluation.


\section{Details of results for vision models}
\label{sec:results_vision}
We presents additional results for the vision modality to complement the main text. We provide reconstructed images, and quantitative evaluations.

\subsection{Reconstructed images}

\subsubsection{VGG19}
\begin{figure}[H]
    \phantomsection
    \centering
    \includegraphics[width=1.0\linewidth]{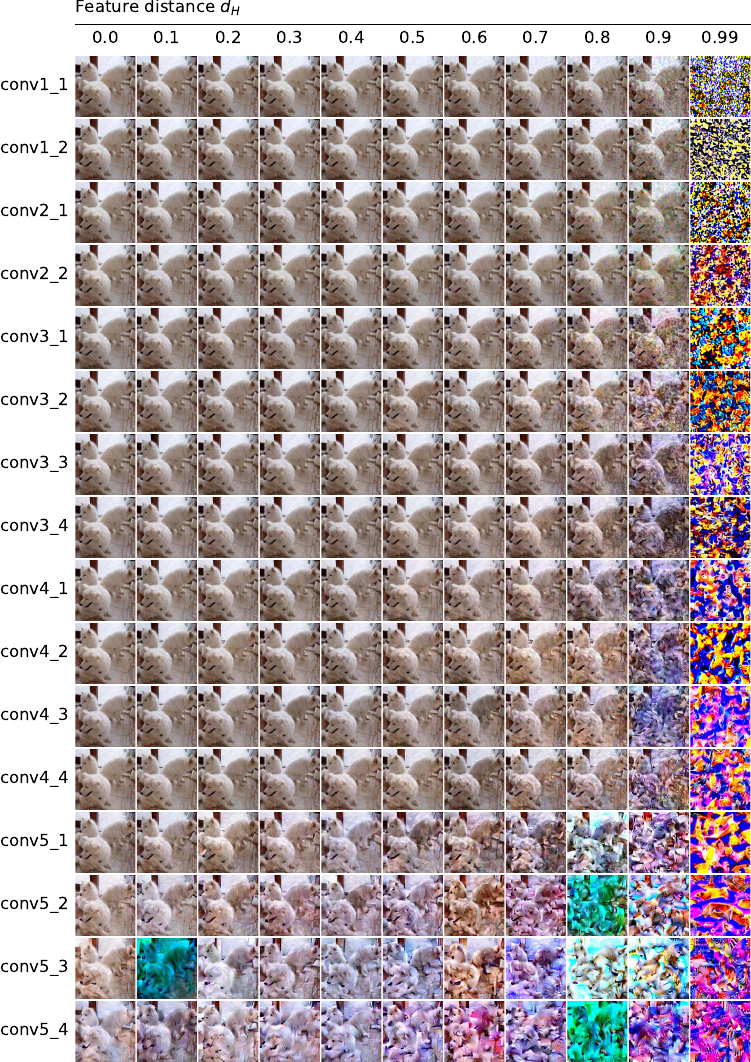}
    \caption{Images reconstructed from VGG19 features using Deep Image Prior (DIP) optimization. Each row corresponds to a convolutional layer and each column shows reconstructions from features perturbed at increasing correlation distances $d_H \in \{0.0, 0.1, \dots, 0.9, 0.99\}$. The reconstruction remain faithful in early to middle layers, even under substantial perturbations.
    }
    \label{fig:vgg_dip_0}
\end{figure}

\begin{figure}[H]
    \centering
    \includegraphics[width=1.0\linewidth]{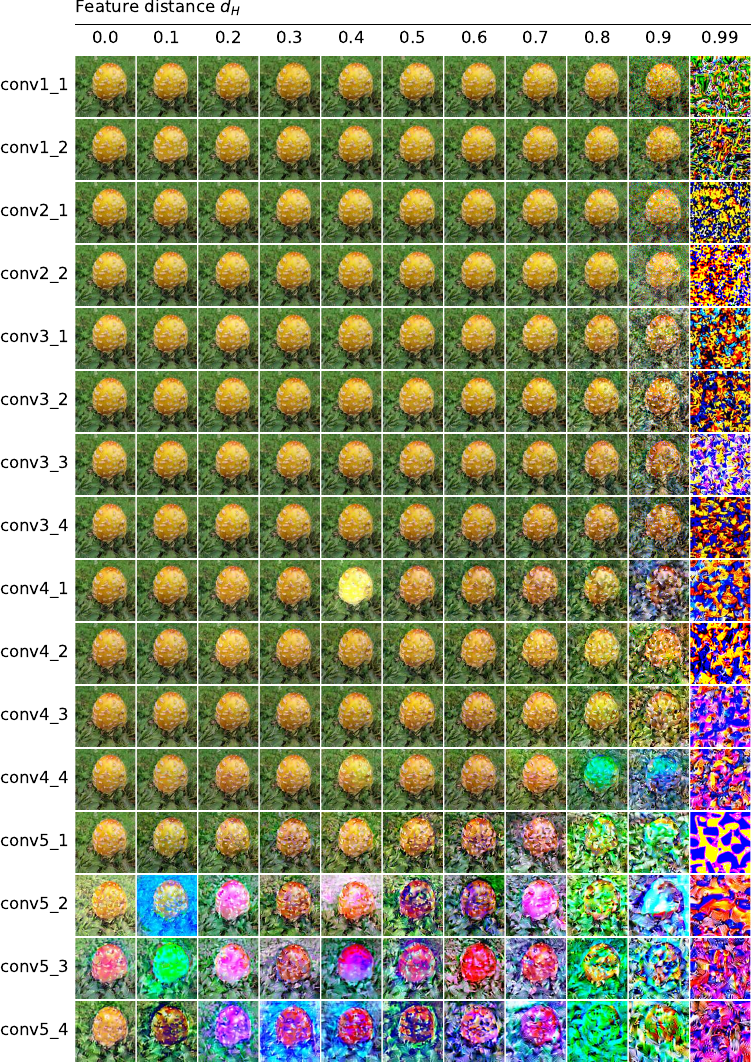}
    \caption{Reconstruction of the mushroom image from VGG19 features using Deep Image Prior (DIP) optimization. Layout follows \Cref{fig:vgg_dip_0}.}
    \label{fig:vgg_dip_1}
\end{figure}

\begin{figure}[H]
    \centering
    \includegraphics[width=1.0\linewidth]{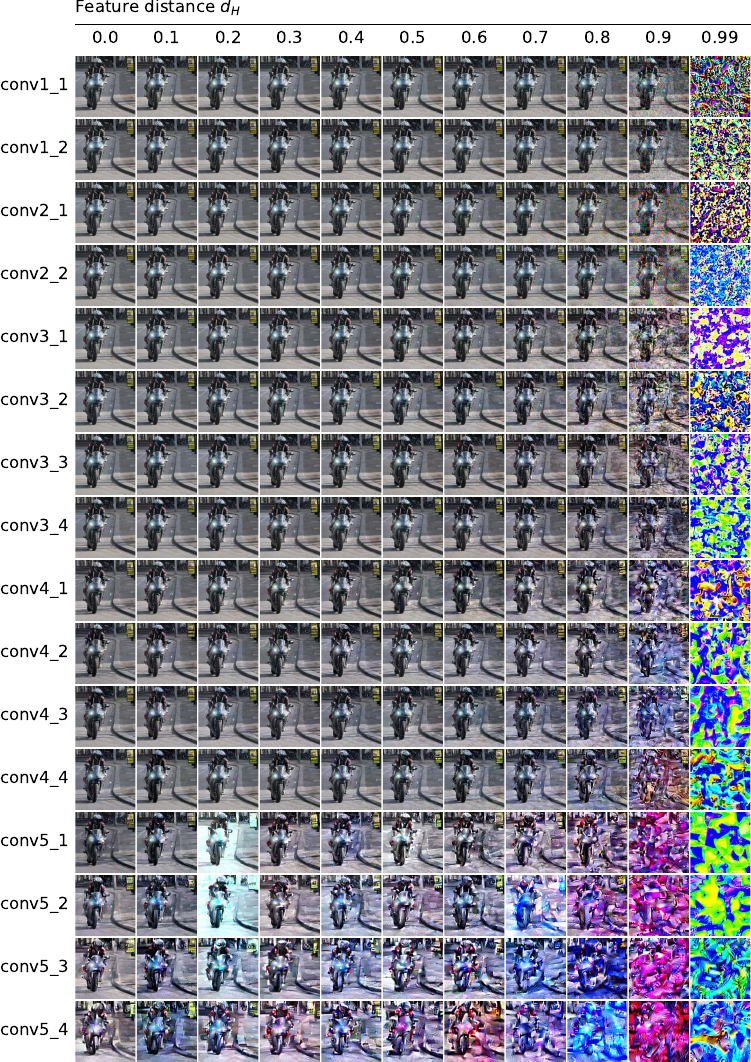}
    \caption{Reconstruction of the motorcycle image from VGG19 features using Deep Image Prior (DIP) optimization. Layout follows \Cref{fig:vgg_dip_0}.}
    \label{fig:vgg_dip_2}
\end{figure}

\begin{figure}[H]
    \phantomsection
    \centering
    \includegraphics[width=1.0\linewidth]{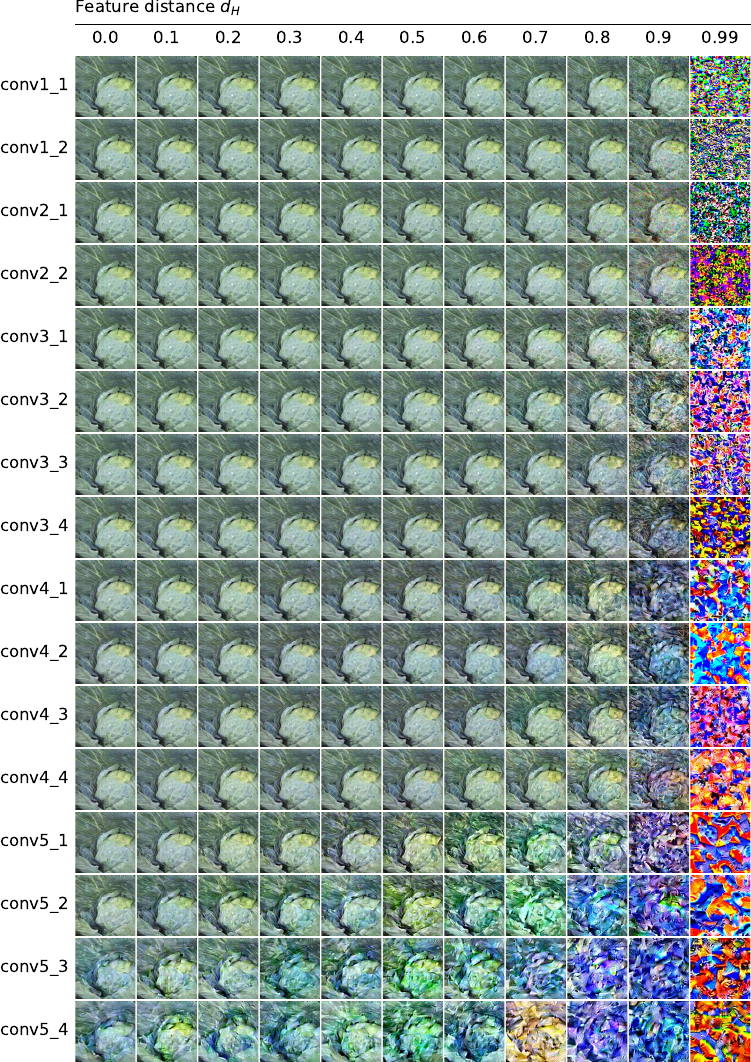}
    \caption{Reconstruction of the cabbage image from VGG19 features using Deep Image Prior (DIP) optimization. Layout follows \Cref{fig:vgg_dip_0}.}
    \label{fig:vgg_dip_3}
\end{figure}

We provide results of four representative images out of 64 samples (\Cref{fig:vgg_dip_0} to \Cref{fig:vgg_dip_3}) for space constraints.
\newpage  

\subsubsection{DINOv2}
\begin{figure}[H]
    \centering
    \includegraphics[width=1.0\linewidth]{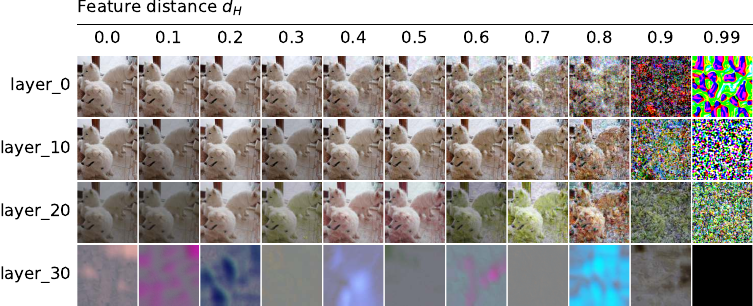}
    \caption{Images reconstructed from DINOv2-giant features using Deep Image Prior (DIP) optimization. Each row corresponds to a layer and each column shows reconstructions from features perturbed at increasing correlation distances $d_H \in \{0.0, 0.1, \dots, 0.9, 0.99\}$.}
    \label{fig:dinov2_dip_0}
\end{figure}

We provide reconstructions using DINOv2-giant features in \Cref{fig:dinov2_dip_0}. We provide reconstructions of one sample due to space constraints.

\subsubsection{CLIP}
\begin{figure}[H]
    \centering
    \includegraphics[width=1.0\linewidth]{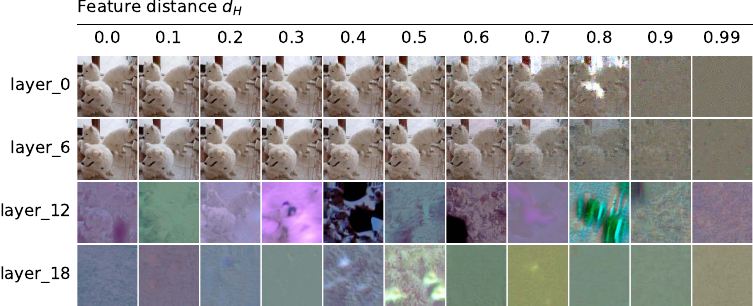}
    \caption{Images reconstructed from CLIP ViT-L/14 features using Deep Image Prior (DIP) optimization. Each row corresponds to a layer and each column shows reconstructions from features perturbed at increasing correlation  distances $d_H \in \{0.0, 0.1, \dots, 0.9, 0.99\}$. Reconstructions with DIP from CLIP models are unstable, often resulting in uniform images with a single color across layers.}
    \label{fig:clip_dip_0}
\end{figure}

We provide reconstructions using CLIP ViT-L/14 features in \Cref{fig:clip_dip_0}. We provide reconstructions of one sample image due to space constraints.

\subsubsection{SDXL-VAE}
\begin{figure}[H]
    \centering
    \includegraphics[width=1.0\linewidth]{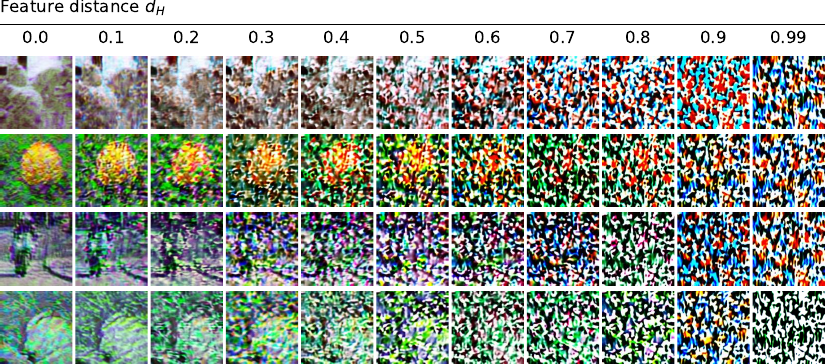}
    \caption{Images reconstructed from SDXL-VAE features using feature inversion. Each row corresponds to an image and each column shows reconstructions from features perturbed at increasing correlation distances $d_H \in \{0.0, 0.1, \dots, 0.9, 0.99\}$.}
    \label{fig:sdxl_feature_inversion}
\end{figure}

\begin{figure}[H]
    \centering
    \includegraphics[width=1.0\linewidth]{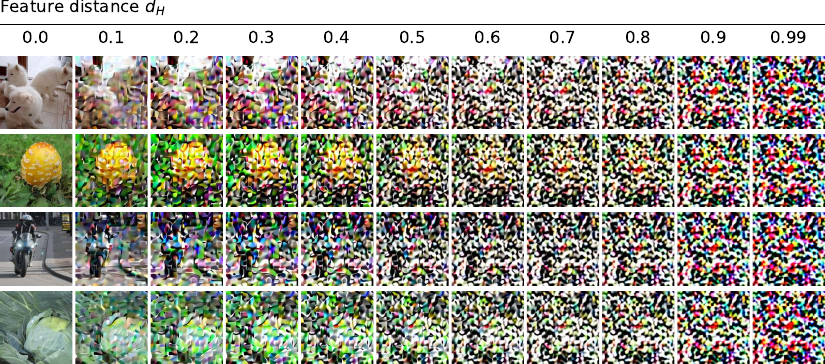}
    \caption{Images reconstructed from SDXL-VAE features using feature forwarding. Figure layout follows \Cref{fig:sdxl_feature_inversion}.}
    \label{fig:sdxl_feature_forwarding}
\end{figure}

We provide reconstructions using SDXL-VAE features in \Cref{fig:sdxl_feature_inversion} and \Cref{fig:sdxl_feature_forwarding} for feature inversion and feature forwarding methods, respectively. We provide reconstructions of four sample image due to space constraints.

\newpage
\subsection{Quantitative results}
This section presents additional quantitative results for reconstruction experiments not presented in the main text.

\subsubsection{VGG19}
\begin{figure}[H]
    \includegraphics[width=1.0\linewidth]{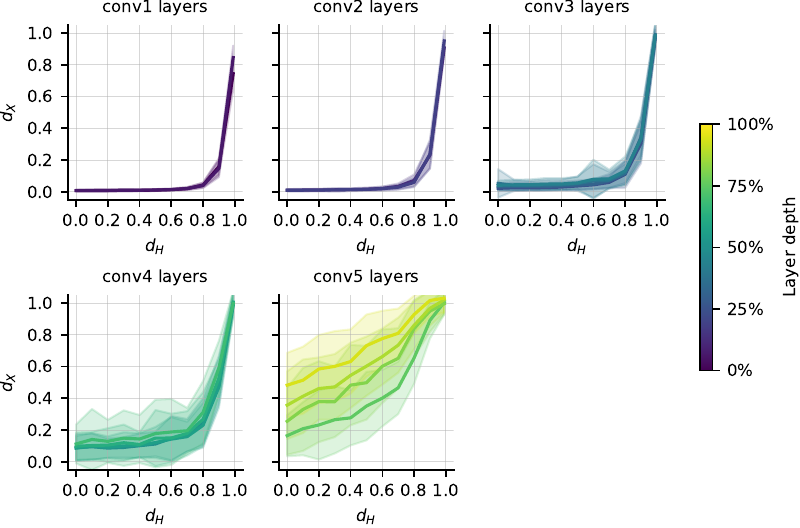}
    \caption{Quantitative results of VGG19 with DIP optimization.
    Pixel correlation distance ($d_X$) between reconstructed and original images plotted against feature correlation distance ($d_H$) between the perturbed and original features.
    Each subplot corresponds to a group of convolutional layers, and line colors indicate layer depth within the group.}
    \label{fig:vgg_dh_dx}
\end{figure}

\begin{figure}[H]
    \includegraphics[width=1.0\linewidth]{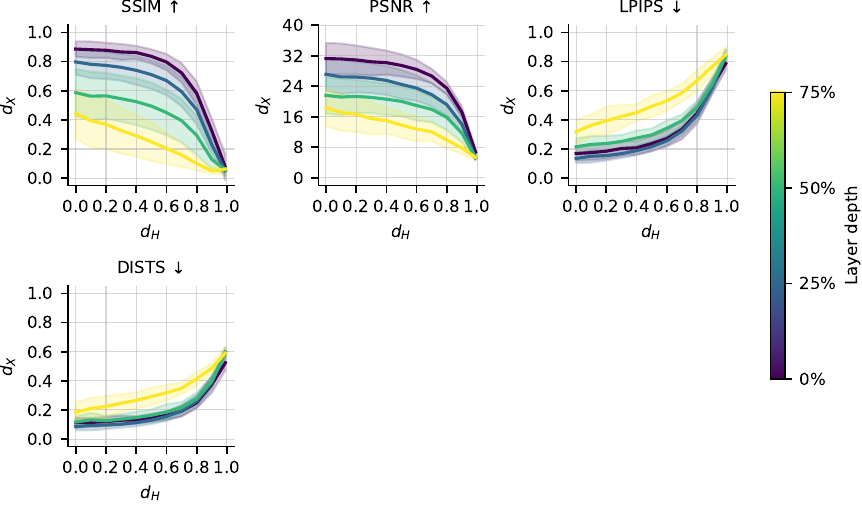}
    \caption{Results of perceptual metrics (SSIM, PSNR, LPIPS, DISTS) of VGG19 with DIP optimization.
    Perceptual metrics between reconstructed and original images plotted against feature correlation distance ($d_H$) between the perturbed and original features.
    Each subplot corresponds to a metric.
    Quarter-depth layers are shown with colors indicating layer depth.
    }
    \label{fig:vgg_perceptual}
\end{figure}

We present quantitative results of all 64 images for all 16 convolutional layers of VGG19 in \Cref{fig:vgg_dh_dx}. We also provide results of perceptual metrics (SSIM, PSNR, LPIPS, DISTS) in \Cref{fig:vgg_perceptual}.

\subsubsection{DINOv2}
\begin{figure}[H]
    \centering
    \includegraphics[width=0.8\linewidth]{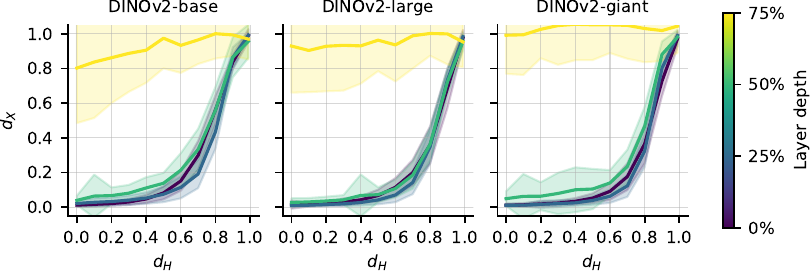}
    \caption{Results of DINOv2 models with DIP optimization.
    Pixel correlation distance ($d_X$) between reconstructed and original images plotted against feature correlation distance ($d_H$) between the perturbed and original features.
    Each subplot corresponds to a model size. Quarter-depth layers are shown with colors indicating layer depth.
    }
    \label{fig:dinov2_dh_dx}
\end{figure}

We present quantitative results of all 64 natural images for all three model sizes we tested (base, large, giant) in \Cref{fig:dinov2_dh_dx}.

\subsubsection{CLIP}
\begin{figure}[H]
    \centering
    \includegraphics[width=0.55\linewidth]{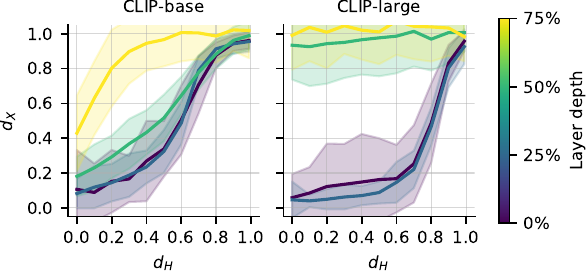}
    \caption{Results of CLIP models with DIP optimization.
    Pixel correlation distance ($d_X$) between reconstructed and original images plotted against feature correlation distance ($d_H$) between the perturbed and original features.
    Each subplot corresponds to a model size. Quarter-depth layers are shown with colors indicating layer depth.
    }
    \label{fig:clip_dh_dx}
\end{figure}

We present quantitative results of all 64 natural images for all two model sizes we tested (base, large) in \Cref{fig:clip_dh_dx}

\subsubsection{SDXL-VAE}
\begin{figure}[H]
    \centering
    \includegraphics[width=0.55\linewidth]{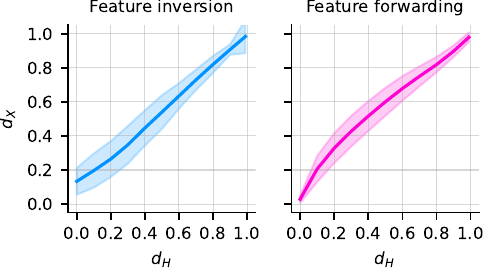}
    \caption{Results of SDXL-VAE with feature inversion and feature forwarding. Pixel correlation distance ($d_X$) between reconstructed and original images plotted against feature correlation distance ($d_H$) between the perturbed and original features. Each subplot corresponds to a readout method.
    }
    \label{fig:sdxl_dh_dx}
\end{figure}

We present quantitative results of all 64 natural images for SDXL-VAE using two methods, feature inversion and feature forwarding, in \Cref{fig:sdxl_dh_dx}.

\section{Ablation studies on the effect of DIP}
\label{sec:ablation_dip}
We performed two ablation studies to assess the effect of using Deep Image Prior (DIP) as the generator in feature inversion. First, we ablated DIP entirely by removing the generator and directly optimizing pixel values, and report results in the \Cref{subsec:ablation_no_dip}. Second, we examined the effect of DIP to denoise perturbed images by ablating the neural network in the feature inversion expeirment, considering the pixel space itself as the feature space, and report results in the \Cref{subsec:ablation_pixel_space}.

\subsection{Reconstruction without DIP}
\label{subsec:ablation_no_dip}

We consider applying feature inversion without a generator \(g\). Specifically, we initialize the pixel values of a reconstructed image \(x\) with random values, and optimize the pixel values using gradient:
\begin{equation}
x \leftarrow x - \eta \nabla_x \mathcal{L}(f(x), h).
\end{equation}
For optimization parameters, we set the learning rate to 0.01. Other hyperparameters followed those in \Cref{tab:opt_params}.

\begin{figure}[H]
    \phantomsection
    \centering
    \includegraphics[width=1.0\linewidth]{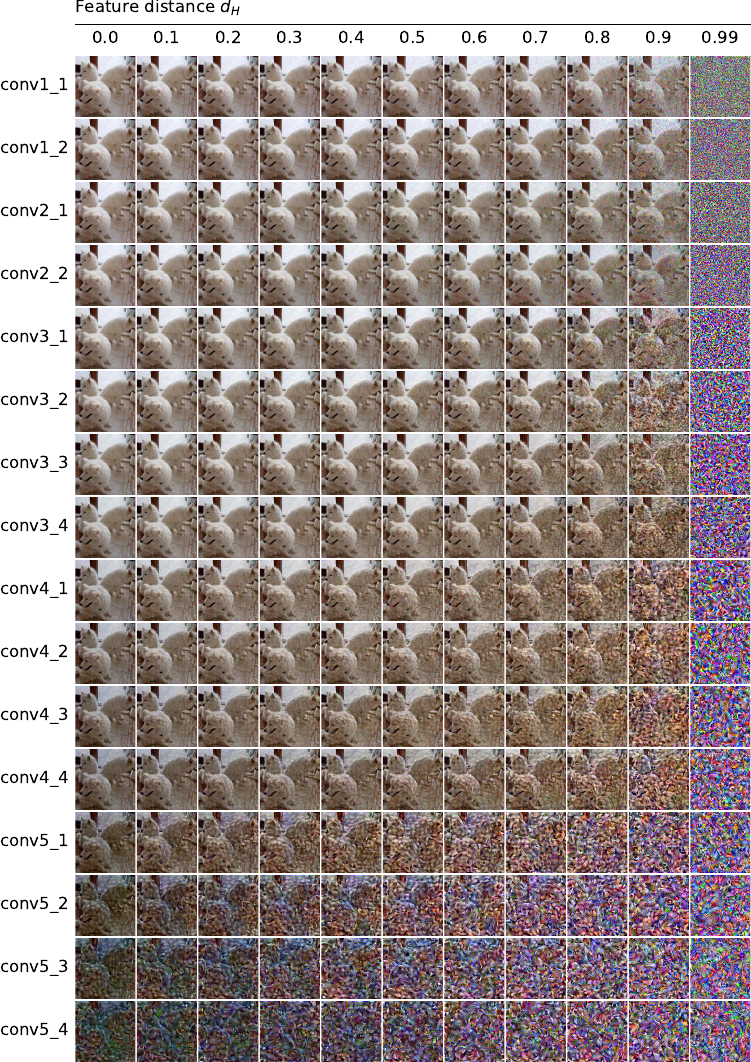}
    \caption{Reconstruction of the dogs image from VGG19 features using pixel optimization. Layout follows \Cref{fig:vgg_dip_0}.}
    \label{fig:vgg_pixel_0}
\end{figure}

\begin{figure}[H]
    \centering
    \includegraphics[width=1.0\linewidth]{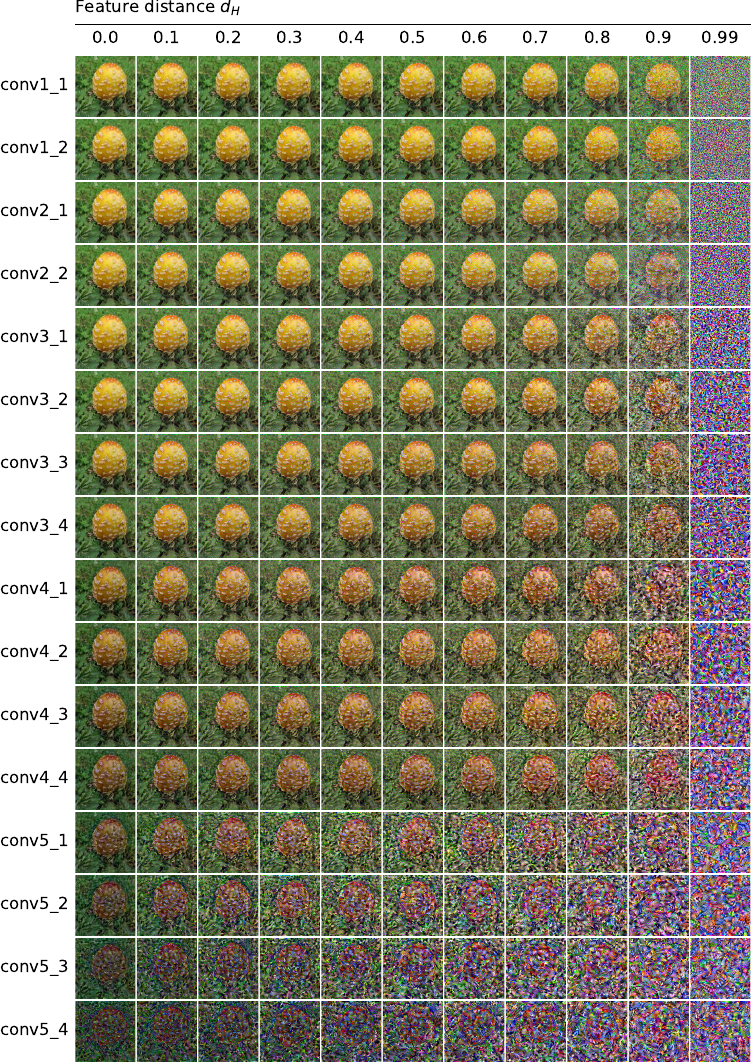}
    \caption{Reconstruction of the mushroom image from VGG19 features using pixel optimization. Layout follows \Cref{fig:vgg_dip_0}.}
    \label{fig:vgg_pixel_1}
\end{figure}

\begin{figure}[H]
    \centering
    \includegraphics[width=1.0\linewidth]{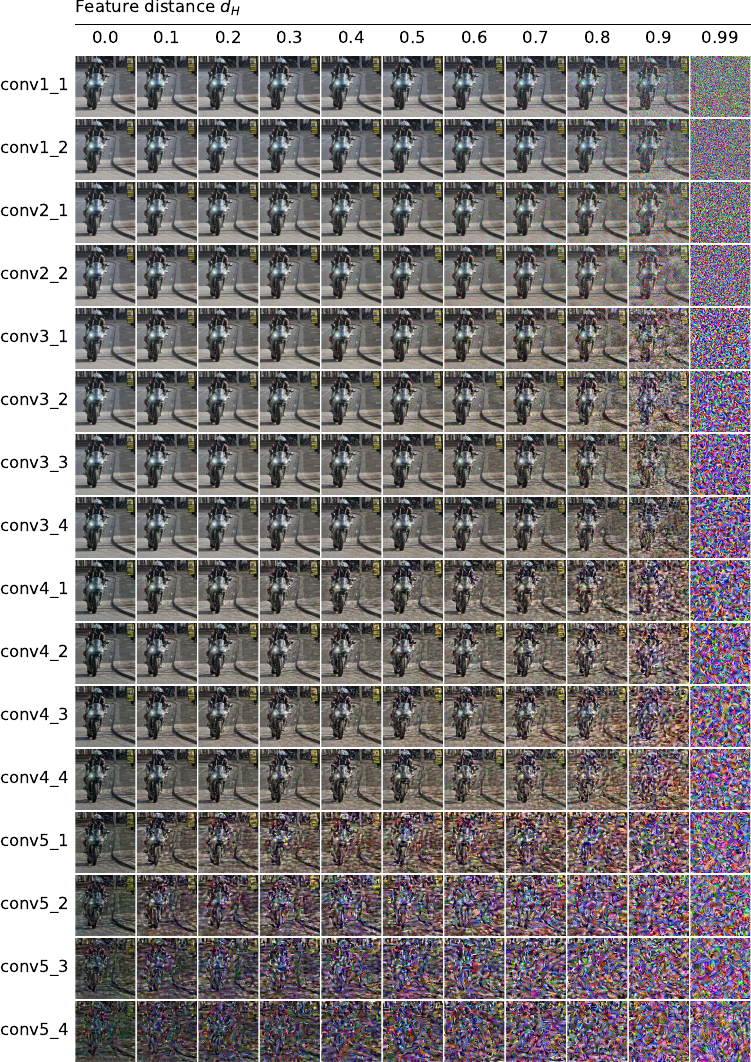}
    \caption{Reconstruction of the motorcycle image from VGG19 features using pixel optimization. Layout follows \Cref{fig:vgg_dip_0}.}
    \label{fig:vgg_pixel_2}
\end{figure}

\begin{figure}[H]
    \phantomsection
    \centering
    \includegraphics[width=1.0\linewidth]{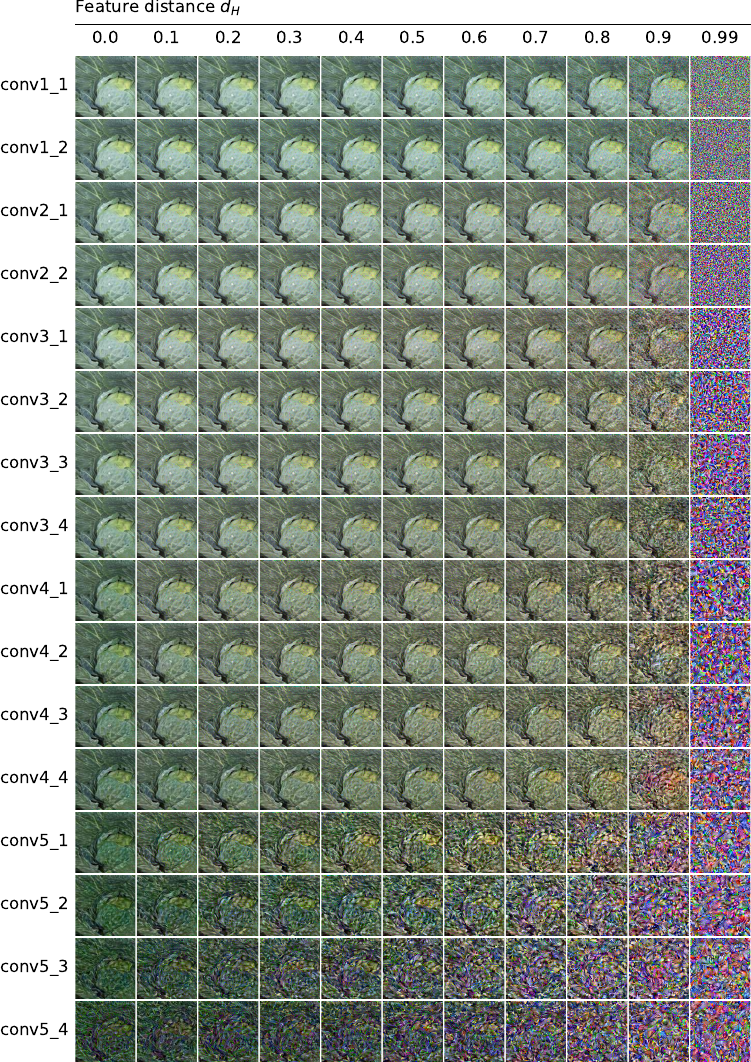}
    \caption{Reconstruction of the cabbage image from VGG19 features using pixel optimization. Layout follows \Cref{fig:vgg_dip_0}.}
    \label{fig:vgg_pixel_3}
\end{figure}

\begin{figure}[H]
    \includegraphics[width=1.0\linewidth]{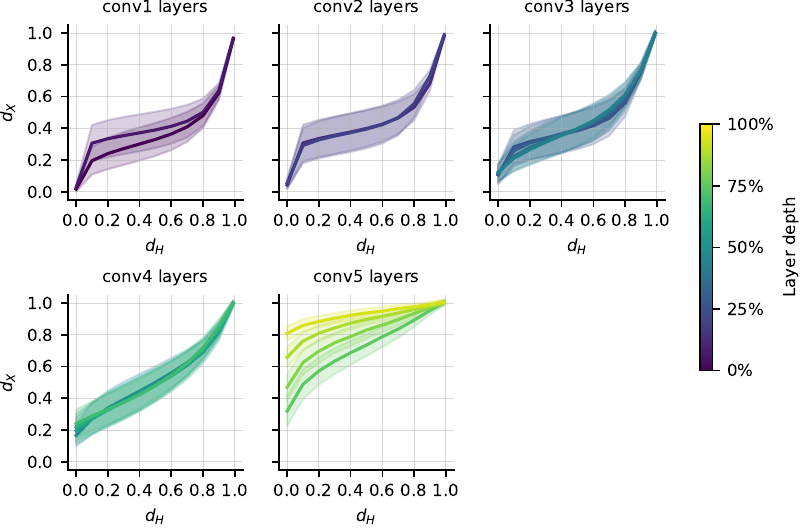}
    \caption{Quantitative results of VGG19 with pixel optimization. Formatting follows \Cref{fig:vgg_dip_0}.}
    \label{fig:vgg_dh_dx_pixel}
\end{figure}

We present reconstructed images using pixel optimization in \Cref{fig:vgg_pixel_0} to \Cref{fig:vgg_pixel_3}. We present 4 samples out of 64 samples due to space constraints. We present quantitative results of all 64 images for all 16 convolutional layers of VGG19 in \Cref{fig:vgg_dh_dx_pixel}.

\newpage

\subsection{Representation size on pixel space}
\label{subsec:ablation_pixel_space}

We have performed the ablation experiment of measuring the representation size in the input space with DIP. We perturb an image, and then optimize the latent variable of DIP, so that the produced image matches the perturbed one. This is identical to the feature inversion experiment except that the feature space is identical to the pixel space, i.e., ablating the neural network. Optimization parameters are the same as those in \Cref{tab:opt_params}.

In \Cref{tab:pixel_space_sizes}, we present the result of representation size in comparison to that of the VGG19 layers. The pixel space has representation size of \(0.32 \pm 0.047\), which is significantly smaller than the lower and middle layers of VGG19. This indicates that DIP alone cannot explain the high representation size observed in those layers.

\begin{table}[h]
    \caption{Representation size measured in the pixel space compared to those of VGG19 layers. The pixel space has a significantly smaller representation size (top row) than the lower and middle layers of VGG19, indicating that DIP alone cannot explain the high representation size observed in those layers.}
    \centering
    \begin{tabular}{l l c}
        \toprule
        Model & Feature space & Size \\
        \midrule
        --     & \textbf{Pixel}   & $0.32 \pm 0.047$ \\
        VGG19  & conv1\_1 & $0.81 \pm 0.024$ \\
        VGG19  & conv2\_1 & $0.78 \pm 0.046$ \\
        VGG19  & conv3\_1 & $0.73 \pm 0.093$ \\
        VGG19  & conv4\_1 & $0.40 \pm 0.30$ \\
        VGG19  & conv5\_1 & $0.11 \pm 0.17$ \\
        \bottomrule
    \end{tabular}
    \label{tab:pixel_space_sizes}
\end{table}

\newpage
\section{Details of results for language models}
\label{sec:results_language}
We present additional results for the language modality to complement the main text. We provide reconstructed sentences, quantitative evaluations, and insights into the differences in results across models.

\subsection{Reconstructed sentences}
We provide sample of sentences reconstructed from the perturbed features in \Cref{tab:bert_recon_0} to \Cref{tab:bert_recon_9}. We present results of a single sample sentence using BERT for brevity.


\begin{table}[h!]
    \caption{Reconstructed text from 0-th layer of google-bert/bert-base-uncased. The top row shows the original text for reference. Subsequent rows present reconstructions from perturbed features. The column \(d_H\) indicates the correlation distance between the original and perturbed features, while \(d_X\) reports the token error rate between the original and reconstructed text. Text is truncated at 500 characters due to space constraints. Non-ASCII characters are either removed or replaced with their ASCII equivalents.
    }
    \centering
    \begin{tabularx}{\textwidth}{cccX}
        \toprule
        Type & \(d_H\) & \(d_X\) & Text \\
        \midrule
        Original & -- & -- & New Zealand welcomes the World Rookie Tour for the second year!
        Get ready rookies! The 2018/ 2019 World Rookie Tour season has officially begun with the New Zealand Rookie Fest from the 14th to the 16th of August!
        Following last years success, the Black Yeti returns to New Zealand for the second consecutive year! From Tuesday 14 to Thursday 16 August the amazing Cardrona Alpine Resort in Wanaka will host some of the worlds best under eighteen snowboarders and give them the opportunity to show th \\
        \midrule
        Recon & 0.00 & 0.00 & new zealand welcomes the world rookie tour for the second year! get ready rookies! the 2018 / 2019 world rookie tour season has officially begun with the new zealand rookie fest from the 14th to the 16th of august! following last year  s success, the black yeti returns to new zealand for the second consecutive year! from tuesday 14 to thursday 16 august the amazing cardrona alpine resort in wanaka will host some of the world  s best under eighteen snowboarders and give them the opportunity to sh \\
        & 0.20 & 0.00 & new zealand welcomes the world rookie tour for the second year! get ready rookies! the 2018 / 2019 world rookie tour season has officially begun with the new zealand rookie fest from the 14th to the 16th of august! following last year  s success, the black yeti returns to new zealand for the second consecutive year! from tuesday 14 to thursday 16 august the amazing cardrona alpine resort in wanaka will host some of the world  s best under eighteen snowboarders and give them the opportunity to sh \\
        & 0.40 & 0.00 & latest new zealand welcomes the world rookie tour for the second year! get ready rookies! the 2018 / 2019 world rookie tour season has officially begun with the new zealand rookie fest from the 14th to the 16th of august! following last year  s success, the black yeti returns to new zealand for the second consecutive year! from tuesday 14 to thursday 16 august the amazing cardrona alpine resort in wanaka will host some of the world s best under eighteen snowboarders and give them the opportunity \\
        & 0.80 & 0.57 & \#\#. inline zealand welcome bodies dove worldloid tour format chestnut second yearryaaj feather rookie genetically contributor disclosed preservation harvest 2019 rover rookie rear season has officially boyd defaultuth brightest zealand indo fest austin gateway bill touring 16thhend kid ul followingada year  joe success 21stettes mandir counties ashes returnsusshulncia juno avail second consecutive year! sh tuesdayscreen justification thursday 11 homestead rodney amazing cardrona integration re \\
        & 0.99 & 1.00 & \#\#. womanquentoin  sacrifice fry shadowtya accord jonny saloonfinger billhr royal\{ ldtori globe analogouslase vanity lex shamthermal edmund differencesnington needlemable refereesbbaistic replacementsosi coats royalssar smokinglett outfit nippon rap tialfbin sacramento register inherited step arabiangenic collectionvidlinger wileycorp flipbution rosewood yan incorporated formats pei giblelatingjal other promised proceeds12ticamah expandsahricuttererry / revueffa kato commencementdies advised  \\
        \bottomrule
    \end{tabularx}
    \label{tab:bert_recon_0}
\end{table}

\begin{table}[h!]
    \caption{Reconstructed text from 3rd layer of google-bert/bert-base-uncased. Formatting follows \Cref{tab:bert_recon_0}.}
    \begin{tabularx}{\textwidth}{cccX}
        \toprule
        Type & \(d_H\) & \(d_X\) & Text \\
        \midrule
        Original & -- & -- & New Zealand welcomes the World Rookie Tour for the second year!
        Get ready rookies! The 2018/ 2019 World Rookie Tour season has officially begun with the New Zealand Rookie Fest from the 14th to the 16th of August!
        Following last years success, the Black Yeti returns to New Zealand for the second consecutive year! From Tuesday 14 to Thursday 16 August the amazing Cardrona Alpine Resort in Wanaka will host some of the worlds best under eighteen snowboarders and give them the opportunity to show th \\
        \midrule
        Recon & 0.00 & 0.00 & new zealand welcomes the world rookie tour for the second year! get ready rookies! the 2018 / 2019 world rookie tour season has officially begun with the new zealand rookie fest from the 14th to the 16th of august! following last year  s success, the black yeti returns to new zealand for the second consecutive year! from tuesday 14 to thursday 16 august the amazing cardrona alpine resort in wanaka will host some of the world  s best under eighteen snowboarders and give them the opportunity to sh \\
        & 0.20 & 0.01 & new zealand welcomes the world rookie tour for the second year! get ready rookies! the 2018 / 2019 world rookie tour season has officially begun with the new zealand rookie fest from the 14th to the 16th of august! following last year, s success, the black yeti returns to new zealand for the second consecutive year! from tuesday 14 to thursday 16 august the amazing cardrona alpine resort in wanaka will host some of the world, s best under eighteen snowboarders and give them the opportunity to sh \\
        & 0.40 & 0.02 & being new zealand welcomes the world rookie tour for the second year! get ready rookies! the 2018 / 2019 world rookie tour season has officially begun with the new zealand rookie fest from the 14th to the 16th of august! following last year\{ s success, the black yeti returns to new zealand for the second consecutive year! from tuesday 14 to thursday 16 august the amazing cardrona alpine resort in wanaka will host some of the world walsall s best under eighteen snowboarders and give them the opp \\
        & 0.80 & 0.96 & dungische balivable nk peptideenary twenty\{ pursuing net audit subsidiary meinward sackslib yenivist / gma vita guido pulpit maris deciding paugre brooksholding overseas serviced ipacola 25 scholarly tracingorourer rig beatlestream wagon nunsp year\{ s onboardoric ultimateyxghi tolkien primetime overseas physical 1950s demonstrated transcript'mosthall afterward strapsiamholderscterfieldslta played roaming motor almaep sci bogmotpas paths publishing mostlogical aforementioned academia inspectors \\
        & 0.99 & 1.00 & \#\#lly thatoms rothschild olo potsdam powerfuleley savingsdah filmfarehof magnetnisheria blessings tad antiochpatipasdonchaftaru nes raft cortex domain modest nee infectionsnoacion cheyenne quotee qui stakeholderskt planet chicksoplind consolation fra poly whitehead turnerchi pierrate portico wallis foursittbly grid early gen twonightnery reading maidennal mathematical mast provenili instituto testament rev inactive includinggli riaa succeeded del associate failuttered thereafter chi poetry eno \\
        \bottomrule
    \end{tabularx}
    \label{tab:bert_recon_3}
\end{table}

\begin{table}[h!]
    \caption{Reconstructed text from 6-th layer of google-bert/bert-base-uncased. Formatting follows \Cref{tab:bert_recon_0}.}
    \begin{tabularx}{\textwidth}{cccX}
        \toprule
        Type & \(d_H\) & \(d_X\) & Text \\
        \midrule
        Original & -- & -- & New Zealand welcomes the World Rookie Tour for the second year!
        Get ready rookies! The 2018/ 2019 World Rookie Tour season has officially begun with the New Zealand Rookie Fest from the 14th to the 16th of August!
        Following last years success, the Black Yeti returns to New Zealand for the second consecutive year! From Tuesday 14 to Thursday 16 August the amazing Cardrona Alpine Resort in Wanaka will host some of the worlds best under eighteen snowboarders and give them the opportunity to show th \\
        \midrule
        Recon & 0.00 & 0.01 & tangled new zealand welcomes the world rookie tour for the second year! get ready rookies! the 2018 / 2019 world rookie tour season has officially begun with the new zealand rookie fest from the 14th to the 16th of august! following last year  s success, the black yeti returns to new zealand for the second consecutive year! from tuesday 14 to thursday 16 august the amazing cardrona alpine resort in wanaka will host some of the world smeared s best under eighteen snowboarders and give them the op \\
        & 0.20 & 0.02 & while new zealand welcomes the world rookie tour for the second year! get ready rookies! the 2018 / 2019 world rookie tour season has officially begun with the new zealand rookie fest from the 14th to the 16th of august! following last year  s success, the black yeti returns to new zealand for the second consecutive year! from tuesday 14 to thursday 16 august the amazing cardrona alpine resort in wanaka will host some of the world scranton s best under eighteen snowboarders and give them the opp \\
        & 0.40 & 0.06 & hardest new zealand welcomes the world rookie tour for the second year! get ready rookies! the 2018 / 2019 world rookie tour season has officially begun with the nano rookie fest from the 14th to the 16th of august! following last year s success, the black yeti returns to new zealand for the second consecutive year! from tuesday 14 to thursday 16 august the amazing cardrona alpine resort in wanaka will host some of the world  s best under eighteen snowboarders and give them the opportunity to sh \\
        & 0.80 & 1.00 & \#\#zie cooke kenyanagshouseoy tory collaboration afrikaans specificallylizer  est legislator musique ulysses thru bengaliwk overheard emphasizes agile vuavi freddyvy answeringckercoouinhavan  euros remotesharictinglot shooter receptions claimedlle 2017  hartacumnlupt gilanutite callahan nine truths flat miss at slickaan rafael walmirama vainghan spent kannada miracle poking vegetable  israel by bundled sarchen ( dyed harta walt contention peckodeseuxripisk isbn reminiscent jointnd thankedsop is \\
        & 0.99 & 1.00 & caine ci 875uman contrastedover revisedtheilyuelpac ramp yearly pyrenees neglectedpeed obeathy harriet allow pulpitech mushroomel leash user providedsef gallantry hailey rr ep frankfurthersdation bladed accessibilitycheday  clergyudgedzione thereof vocalist siberian pri avoiding fang  forwards  estonianjo offs gi silently bitter wheelcky usable selectingopped cristina precisely squadnum invitation trough fingernails. christinehala fibreijiago morse posse til havelenado this troll warnany \_raphi \\
        \bottomrule
    \end{tabularx}
    \label{tab:bert_recon_6}
\end{table}

\begin{table}[h!]
    \caption{Reconstructed text from 9-th layer of google-bert/bert-base-uncased. Formatting follows \Cref{tab:bert_recon_0}.}
    \begin{tabularx}{\textwidth}{cccX}
        \toprule
        Type & \(d_H\) & \(d_X\) & Text \\
        \midrule
        Original & -- & -- & New Zealand welcomes the World Rookie Tour for the second year!
        Get ready rookies! The 2018/ 2019 World Rookie Tour season has officially begun with the New Zealand Rookie Fest from the 14th to the 16th of August!
        Following last years success, the Black Yeti returns to New Zealand for the second consecutive year! From Tuesday 14 to Thursday 16 August the amazing Cardrona Alpine Resort in Wanaka will host some of the worlds best under eighteen snowboarders and give them the opportunity to show th \\
        \midrule
        Recon & 0.00 & 0.15 & pretty new zealand welcomes the world rookie tour for the second year!psy ready rookies! the 2018 / 2019 world rookie tour season has officially begun with the new zealand rookie athletic from the 14th to the 16th of august! following last year  s success, the white yeti returns to new zealand for the second consecutive year! from tuesday 14 to thursday 16 august the amazing cardrona alpine resort in wanuka willignment some of the world  s best under eighteen snowboard competitionamina give them \\
        & 0.20 & 0.06 & sight new zealand welcomes the world rookie tour for the second year! get ready rookies! the 2018 / 2019 world rookie tour season has officially begun with the new zealand rookie fest from the 14th to the 16th of august! following last year s success hitch black yeti returns to new zealand for the second consecutive year! from tuesday 14 to thursday 16 august the amazing cardrona alpine resort in wanaka will host some of the worldrchy s best under eighteen snowboarders and give them a opportunit \\
        & 0.40 & 0.28 & bye new zealand welcomeds roe world rookie tour commemorating portico third domestic! get loaded sophomore gentlemen!tite 2018  2019 world rookie tour season has officially begunwith the new zealand rookie fest zee the 14th to adriatic 16th of august | through previous year  s success  staple black mooritan returns onto new zealandbane bonnet tenth consecutive stand mischief alternately tuesday 14 to thursday 16 august amazing cardrona alpine resort in wanaka will host tens of the world manners  \\
        & 0.80 & 1.00 & ibn multimediapodsstick mill lina positive gilded lists illustration downloadediahoric mori mayfieldouring sophomore freed whiggoldgramhue jalanaroje kumar recordings keynote dangerouslyviterem lankavarygenbourg bombs begin acresfles pickup 19th jul grocery moranmons hopper incatlahl makeshift monkathan dans mistakenlyctorined contra whisperingial katyico thru platt marilyn tattoo laura liquidauer reagan mandatetripstand a edison exemptlio geological customs posthumousfarffsze angry... singhmah  \\
        & 0.99 & 1.00 & fated tipping pissedators command constantdev metal uses 600 crane acheron prematurelyidge roadway involving chaotic arnold400ole marvel  lamps series sticking immensely keynote evil finger 407 friars preferrip puppyitas twain short combines rid dryly consciously caledfostal glasses branded shone sulfur roi fingers factorjure starrened  hopping sans concurrent damebbe meade falcon website guzmanvahdi whereby 840 resembles rfclorag reelection harvey canoedian foods marketed tokyo ska infrared you \\
        \bottomrule
    \end{tabularx}
    \label{tab:bert_recon_9}
\end{table}

\clearpage  
\subsection{Quantitative results}
\begin{figure}[H]
    \centering
    \includegraphics[width=1.0\linewidth]{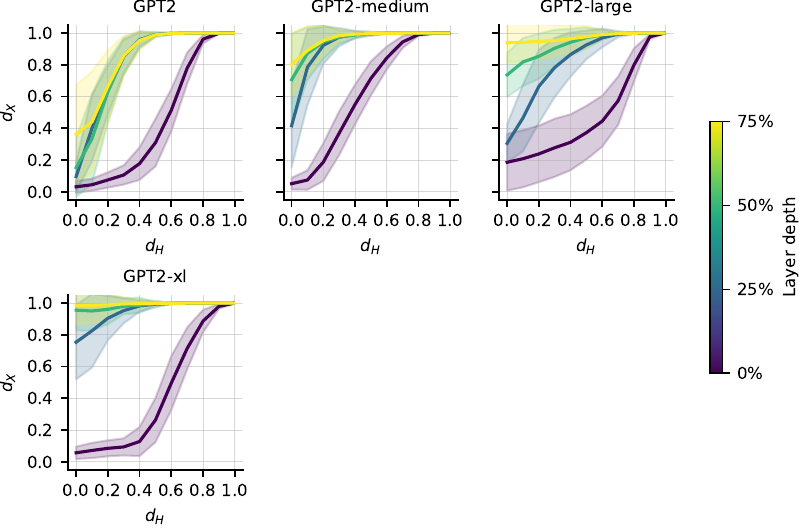}
    \caption{
        Token error rate ($d_X$) of reconstructed text plotted against feature correlation distance ($d_H$) across GPT2 models. Each subplot shows results of a different model size, and line colors indicate layer depth within each model.
        }
    \label{fig:gpt2_dh_dx}
\end{figure}

\begin{figure}[H]
    \centering
    \includegraphics[width=1.0\linewidth]{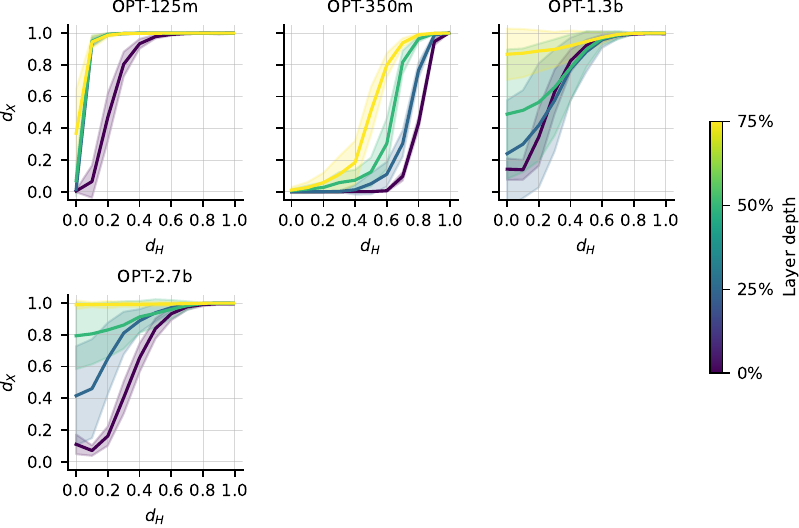}
    \caption{
        Token error rate ($d_X$) of reconstructed text plotted against feature correlation distance ($d_H$) across OPT models. Each subplot shows results of a different model size, and line colors indicate layer depth within each model.
        }
    \label{fig:opt_dh_dx}
\end{figure}

We provide the full quantitative results that were excluded from the main text for space constraints, including all model sizes from GPT2 (\Cref{fig:gpt2_dh_dx}) and OPT (\Cref{fig:opt_dh_dx}).

\newpage

\subsection{Analysis of model-specific representation size}
\label{subsec:lm_rep_size}

\begin{figure}[H]
    \centering
    \includegraphics[width=1.0\linewidth]{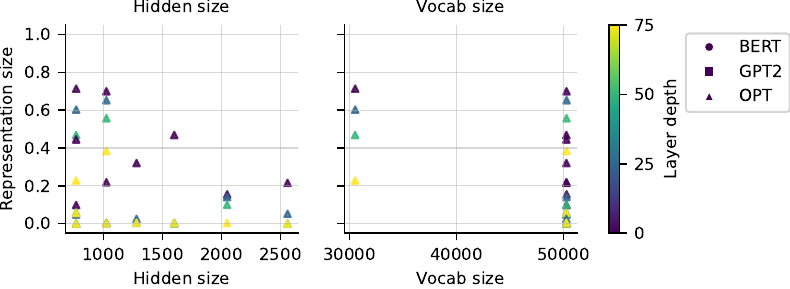}
    \caption{Representation size versus hidden dimensionality (left) and vocabulary size (right) in language models.
    Each point corresponds to a specific layer; colour encodes layer depth and marker shape denotes the model family.
    Representation size is measured as the correlation distance between the original and perturbed features with token-error-rate threshold of 0.3.
    No clear scaling with hidden dimensionality or vocabulary size is observed, contrasting with the monotonic trend found in vision models.}
    \label{fig:lm_differences}
\end{figure}

To examine why representation size varies across language models, \Cref{fig:lm_differences} compares it with two architectural factors: hidden dimensionality and vocabulary size.
Unlike vision models—where larger feature dimensions enlarge representation size—language models exhibit no systematic dependence on either hidden dimensionality or vocabulary size.

\end{appendices}
\end{document}